\documentclass[preprint,amsmath,amssymb,aps,prd,nofootinbib]{revtex4}
\usepackage{epsfig,graphicx,color}
\begin{document}
\begin{flushright}
KIAS-P12034
\end{flushright}
\title{
Recent Neutrino Data and Type III Seesaw \\
with Discrete Symmetry
}

\author{Y. H. Ahn$^1$\footnote{Email: yhahn@kias.re.kr},
C. S. Kim$^2$\footnote{Email: cskim@yonsei.ac.kr},
and Sechul Oh$^3$\footnote{Email: scohph@yonsei.ac.kr}}

\affiliation{
$^1$School of Physics, KIAS, Seoul 130-722, Korea  \\
$^2$Department of Physics, Yonsei University, Seoul 120-749, Korea \\
$^3$University College, Yonsei University, Incheon 406-840, Korea
}

\date{\today}

\begin{abstract}
\noindent
In light of the recent neutrino experiment results from Daya Bay and RENO Collaborations,
we study phenomenology of neutrino mixing angles in the Type III seesaw model with an discrete $A_4 \times Z_2$
symmetry, whose spontaneously breaking scale is much higher than the electroweak scale.
At tree level, the tri-bimaximal (TBM) form of the lepton mixing matrix can be obtained from leptonic
Yukawa interactions in a natural way.
We introduce all possible effective dimension-5 operators, invariant under the Standard Model
gauge group and $A_4 \times Z_2$, and explicitly show that they induce a deviation of the lepton mixing
from the TBM mixing matrix,  which can explain a large mixing angle $\theta_{13}$ together with small deviations
of the solar and atmospheric mixing angles from the TBM.
Two possible scenarios are investigated, by taking into account either negligible or sizable
contributions from the light charged lepton sector to the lepton mixing matrix.
Especially it is found in the latter scenario that all the neutrino experimental data, including the recent
best-fit value of $\theta_{13} = 8.68^{\circ}$, can be accommodated. The leptonic $CP$ violation characterized by the
Jarlskog invariant $J_{CP}$ has a non-vanishing value, indicating a signal of maximal $CP$ violation.
\end{abstract}

\maketitle %
\section{Introduction}

Recent analyses on the knowledge of neutrino oscillation parameters make desirable a neutrino texture
going beyond the mere fitting procedure~\cite{valle,Reactor,Machado:2011ar}, indicating that neutrinos are massive and leptons
of different families mix with each other in the charged weak interaction. The recent measurements of the leptonic mixing angle $\theta_{13}$
by Daya Bay and RENO Collaborations~\cite{Reactor} indicate that the tri-bimaximal mixing (TBM)~\cite{Harrison:2002er},
giving $\sin^2 \theta_{12} = 1/3$, $\sin^2 \theta_{23} = 1/2$ and $\sin\theta_{13} = 0$, should be modified.
This result is in good agreement with the previous data from T2K, MINOS and Double Chooz Collaborations~\cite{Reactor}, and Daya Bay and RENO progresses have led us to accomplish the measurements of three mixing angles,
$\theta_{12}, \theta_{23}$ and $\theta_{13}$ from three kinds of neutrino oscillation experiments.
A combined analysis of the data coming from T2K, MINOS, Double Chooz and Daya Bay experiments shows~\cite{Machado:2011ar} that
\begin{eqnarray}
\sin^2 2\theta_{13}=0.089\pm 0.016(0.047)~,
\end{eqnarray}
or equivalently,
\begin{eqnarray}
\theta_{13}=8.68^{\circ+0.77^{\circ}~(+2.14^{\circ})}_{~-0.84^{\circ}~(-2.76^{\circ})}
 \label{expdata}
\end{eqnarray}
at $1\sigma~(3\sigma)$ levels and that the hypothesis $\theta_{13}=0$ is now rejected at a significance level higher than $6\sigma$. Although neutrinos have gradually revealed their properties in various experiments since the historic
Super-Kamiokande confirmation of neutrino oscillations~\cite{Fukuda:1998mi}, properties related to the
leptonic CP violation are completely unknown yet. In addition, the large values of the solar mixing
angle $\theta_{\rm sol} \simeq \theta_{12}$ and the atmospheric mixing angle
$\theta_{\rm atm} \simeq \theta_{23}$ may be telling us about some new symmetries of leptons not
presenting in the quark sector and may provide a clue of the nature in quark-lepton physics beyond the
standard model (SM).

The $\mu-\tau$ symmetry, which is the most popular discrete symmetry, has made some success in
describing the masses and mixing pattern in the lepton sector~\cite{mutau}.
Furthermore, Ma and Rajasekaran~\cite{Ma:2001dn} have introduced for the first time the $A_{4}$ flavor
symmetry to avoid the mass degeneracy between $\mu$ and $\tau$ under the $\mu-\tau$ symmetry.
In a well-motivated extension of the SM with the $A_{4}$ symmetry~\cite{He:2006dk}, the TBM pattern of
the lepton mixing matrix comes out in a natural way.
Models with the $A_4$ symmetry combined with grand unification~\cite{Altarelli:2008bg},
supersymmetry~\cite{Bazzocchi:2007na} and extra dimensions~\cite{Altarelli:2005yp,Altarelli:2006kg}
have been also investigated extensively in the literature.

On the other hand, among many possibilities proposed to understand the tiny masses of neutrinos,
the most popular are the seesaw scenarios in which the light neutrino masses become small due to
sufficiently large masses of newly introduced particles.
There are three different types of the seesaw models:

$\bullet$ Type I seesaw with three heavy right-handed Majorana neutrinos~\cite{type1_seesaw},

$\bullet$ Type II seesaw where an electroweak Higgs triplet is used to directly provide the left-handed
neutrinos with small Majorana masses~\cite{type2_seesaw},

$\bullet$ Type III seesaw introducing $SU(2)_L$ fermion triplets with zero hypercharge~\cite{type3_seesaw}.

The Type I and Type II seesaw models with the $A_4$ flavor symmetry (and an auxiliary symmetry) have
been extensively studied in the literature~\cite{He:2006dk, other}.
In this work, we carry out a systematic study of neutrino phenomenology in the Type III
seesaw model with the $A_4$ symmetry, which is spontaneously broken at a scale much higher than the
electroweak scale.  The fermion triplet in the Type III seesaw model transforms under the SM gauge
group $SU(3)_C\times SU(2)_L\times U(1)_Y$ as (1,3,0).
We assume that there are three copies of such fermion triplets.
Among many interesting features~\cite{model} of the model are the possibility of having low seesaw
scale of order a TeV to realize leptogenesis~\cite{leptogenesisIII} and detectable effects at
LHC~\cite{production} through gauge interactions of the heavy triplet leptons or through relatively
large mixing of the light and heavy neutrinos, and the possibility of having new tree level FCNC
interactions in the lepton sector~\cite{fcnc}.

By combining the $A_4$ flavor symmetry with the seesaw mechanism embedded in the Type III model, we
show that the TBM pattern of the lepton mixing matrix as well as the tiny neutrino masses can be
understood at tree level in our framework.  We further investigate the possibility that all the neutrino
experimental data can be accommodated in our framework through the effects from higher dimensional
operators.
For this goal we introduce all possible effective dimension-5 operators, invariant under
$SU(2)_{\rm L} \times U(1)_{\rm Y} \times A_{4} \times Z_{2}$, both in the neutrino
and in the charged lepton sector.
These dimension-5 operators generate the necessary off-diagonal elements of each mixing matrix induced,
respectively, from the neutrino and charged lepton sectors. Subsequently a deviation of the lepton
mixing matrix from the TBM form is induced so that the non-zero mixing angle
$\theta_{13}$~\cite{Parattu:2010cy} and small deviations from TBM of solar and atmospheric mixing angles can be explained through phase effects~\cite{Ahn:2011yj}.

\section{Type III seesaw with $A_4 \times Z_2$ symmetry $-$ Tri-bimaximal mixing}

In the Type I seesaw model, the seesaw mechanism is realized by introducing heavy right-handed
Majorana neutrinos ($N_R$) that are singlets under the SM gauge groups~\cite{type1_seesaw}.
In the Type III seesaw, the heavy Majorana neutrinos in the Type I seesaw are replaced by SU(2)$_L$
triplets of heavy right-handed leptons having zero hypercharge~\cite{type3_seesaw}.
The component fields of the right-handed triplet $\Sigma$ and the corresponding left-handed one
$\tilde \Sigma^c$ are
\begin{eqnarray}
 \Sigma = {\left(\begin{array}{cc}
  N_{R}/\sqrt{2} & E^{+}_{R} \\
  E^{-}_{R} & -N_{R}/\sqrt{2}
 \end{array}\right)}~, ~~~
 \tilde \Sigma^{c} = {\left(\begin{array}{cc}
  (N_{R})^c /\sqrt{2} & (E^{-}_{R})^c \\
  (E^{+}_{R})^c & -(N_{R})^c /\sqrt{2}
 \end{array}\right)}~,
\end{eqnarray}
where $\tilde \Sigma^{c} \equiv (i \tau_2) \Sigma^c (i \tau_2)$ with the charge conjugate
$\Sigma^c = C \bar \Sigma^T$ and  $\tau_{2}$ the Pauli matrix~\cite{Bandyopadhyay:2009xa}.

Unless flavor symmetries are assumed, particle masses and mixings are generally undetermined in gauge
theory.
To understand the present neutrino oscillation data, we consider $A_{4}$ flavor symmetry together with
an auxiliary symmetry $Z_2$ for leptons.  Then the symmetry group for the lepton sector is
$SU(2)_{L}\times U(1)_{Y} \times A_4 \times Z_2$.
To impose the $A_4$ flavor symmetry on our models properly, the Higgs field sector is extended by
introducing two types of new scalar fields, $\chi$ and $\eta$, besides the usual
SM Higgs field $\Phi$.  The $\chi$ is a $SU(2)_{L}$ singlet and electrically neutral,
but the $\eta$ is a $SU(2)_{L}$ doublet such as $\Phi$:
\begin{eqnarray}
  \Phi =
  {\left(\begin{array}{c}
  \varphi^{+} \\
  \varphi^{0}
 \end{array}\right)}~, ~~~\chi^0 ~,
 ~~~\eta =
  {\left(\begin{array}{c}
  \eta^{+} \\
  \eta^{0}
 \end{array}\right)}~.
  \label{Higgs}
\end{eqnarray}

The field assignments under $SU(2)_L \times U(1)_Y \times A_{4}\times Z_{2}$ in our models are shown
in Table~\ref{reps}, where $L_{L}=(\nu_{L},\ell^{-}_{L})^T$~ is the SM lepton doublet.
Here we recall that $A_{4}$ is the symmetry group of the tetrahedron, or equivalently, the finite group
of the even permutation of four objects.  It has four irreducible representations: one
three-dimensional representation $({\bf 3})$ and three inequivalent one-dimensional representations
$({\bf 1},~ {\bf 1}',~ {\bf 1}'')$.  Their multiplication rules are
~${\bf 3}\otimes{\bf 3} = {\bf 3}_{s}\oplus{\bf 3}_{a}\oplus{\bf 1}\oplus{\bf 1}'\oplus{\bf 1}''$,
~${\bf 1}'\otimes{\bf 1}'' = {\bf 1}$, ${\bf 1}'\otimes{\bf 1}'={\bf 1}''$~ and
~${\bf 1}''\otimes{\bf 1}''={\bf 1}'$.
By denoting two $A_4$ triplets as $a = (a_{1}, a_{2}, a_{3})$ and $b = (b_{1}, b_{2}, b_{3})$, one
obtains
\begin{eqnarray}
 (a\otimes b)_{{\bf 3}_{\rm s}} &=& (a_{2}b_{3}+a_{3}b_{2}, a_{3}b_{1}+a_{1}b_{3}, a_{1}b_{2}+a_{2}b_{1})~,
  \nonumber\\
 (a\otimes b)_{{\bf 3}_{\rm a}} &=& (a_{2}b_{3}-a_{3}b_{2}, a_{3}b_{1}-a_{1}b_{3}, a_{1}b_{2}-a_{2}b_{1})~,
  \nonumber\\
 (a\otimes b)_{{\bf 1}} &=& a_{1}b_{1}+a_{2}b_{2}+a_{3}b_{3}~,\nonumber\\
 (a\otimes b)_{{\bf 1}'} &=& a_{1}b_{1}+\omega a_{2}b_{2}+\omega^{2}a_{3}b_{3}~,\nonumber\\
 (a\otimes b)_{{\bf 1}''} &=& a_{1}b_{1}+\omega^{2} a_{2}b_{2}+\omega a_{3}b_{3}~,
\end{eqnarray}
where $\omega=e^{i2\pi/3}$ is a complex cubic-root of unity.

\begin{table}[t]
\caption{\label{reps} Representations of the fields under $SU(2)_L \times U(1)_Y \times A_4 \times Z_2$.}
\begin{ruledtabular}
\begin{tabular}{cccccccccccc}
Field & $L_{L}$ & $l_R,l'_R,l''_R$ & $\Sigma$ & $\Phi$ & $\eta$ & $\chi$ \\
\hline
$A_4$ & $\mathbf{3}$ & $\mathbf{1}$, $\mathbf{1^\prime}$, $\mathbf{1^{\prime\prime}}$ & $\mathbf{3}$ &
$\mathbf{3}$ & $\mathbf{1}$ & $\mathbf{3}$ \\
$Z_2$ & $+$ & $+$ & $-$ & $+$ & $-$ & $+$ \\
$SU(2)_L\times U(1)_Y$ & $(2,-1)$ & $(1,-2)$ & $(3,0)$ & $(2,1)$ & $(2,1)$ & $(1,0)$ \\
\end{tabular}
\end{ruledtabular}
\end{table}

The $SU(2)_L \times U(1)_Y \times A_4 \times Z_2$ invariant Yukawa Lagrangian for the lepton sector
can be expressed as
\begin{eqnarray}
 -{\cal L}_{\rm Yuk} &=& y_{\Sigma} (\overline{L_L} \Sigma)_{{\bf 1}} ~\tilde{\eta}
 + \frac{1}{2} M ~{\rm Tr}[(\overline{\tilde \Sigma^{c}}\Sigma)_{{\bf 1}}]
 + \frac{1}{2} \lambda^{s}_{\chi} ~{\rm Tr}[(\overline{\tilde \Sigma^{c}}\Sigma)_{{\bf 3}_{s}}] \cdot \chi
 + \frac{1}{2} \lambda^{a}_{\chi} ~{\rm Tr}[(\overline{\tilde \Sigma^{c}}\Sigma)_{{\bf 3}_{a}}] \cdot \chi
 \nonumber\\
 &&+ y_{e} (\overline{L_L} \Phi)_{{\bf 1}} \ell_{R}
 + y_{\mu} (\overline{L_L} \Phi)_{{\bf 1}'} \ell''_{R}
 + y_{\tau} (\overline{L_L} \Phi)_{{\bf 1}''} \ell'_{R} ~+ {\rm H.c.}~,
\label{L_Yuk}
\end{eqnarray}
where ~$\tilde{\eta} \equiv i\tau_{2} ~\eta^{\ast}$.
In the above Lagrangian, the SM charged lepton sector has three independent Yukawa terms with the
couplings ~$y_e, ~y_{\mu}$~ and ~$y_{\tau}$,~ respectively, all involving the $A_4$ triplet Higgs field
~$\Phi$.  The neutrino Dirac term arises from ~$(\overline{L_L} \Sigma)_{{\bf 1}} ~\tilde{\eta}$,~ which
involves only one Yukawa coupling ~$y_{\Sigma}$~ and the $A_4$ singlet ~$\tilde \eta$.~
The right-handed Majorana neutrino terms are associated with a bare mass $M$ and an SM gauge singlet scalar
field $\chi$ which is a $A_4$ triplet.
We will see later that the ${\bf 3_a}$ term ~${\rm Tr}[(\overline{\tilde \Sigma^{c}}\Sigma)_{{\bf 3}_{a}}]
\cdot \chi$~ turns out to give no contributions.
By imposing the additional symmetry $Z_{2}$ as shown in Table~\ref{reps}, the
$SU(2)_L \times U(1)_Y \times A_4$ invariant Yukawa term ~$\overline{L_L} \Sigma ~\Phi$~ is forbidden
from the Lagrangian.

We assume that the vacuum expectation values (VEVs) of the $A_4$ triplet $\Phi$ can be equally aligned,
$i.e.$, $\langle \varphi^{0}\rangle=(v, v, v)$.  The mass matrix $m_{\ell}$ of the SM charged leptons is
derived from the terms associated with the three Yukawa couplings ~$y_e, ~y_{\mu}, ~y_{\tau}$~ as
\begin{eqnarray}
 m_{\ell}= U_{\omega}
 {\left(\begin{array}{ccc}
 \sqrt{3}y_{e}\upsilon &  0 &  0 \\
 0 &  \sqrt{3}y_{\mu}\upsilon &  0 \\
 0 &  0 &  \sqrt{3}y_{\tau}\upsilon
 \end{array}\right)}~,~~~~~~~~~~{\rm with}~~
 U_{\omega}=\frac{1}{\sqrt{3}}{\left(\begin{array}{ccc}
 1 &  1 &  1 \\
 1 &  \omega &  \omega^{2} \\
 1 &  \omega^{2} &  \omega
 \end{array}\right)} ~.
\label{m_ell}
\end{eqnarray}
The above form of ~$m_\ell$~ indicates that the left- and the right-diagonalization matrices,
$U_L^{\ell}$ and $U_R^{\ell}$, for the SM charged lepton sector are identical to $U_{\omega}$ and the
$3 \times 3$ identity matrix $I$, respectively: $i.e.$, the diagonal mass matrix $\hat m_{\ell}$ of the
SM charged leptons is given by
\begin{eqnarray}
 \hat m_{\ell} = (U_L^{\ell})^{\dagger} ~m_{\ell} ~U_{R}^{\ell}
 = U_{\omega}^{\dag} ~m_{\ell} = \sqrt{3}v ~{\rm diag}(y_{e}, y_{\mu}, y_{\tau})
 \equiv {\rm diag}(m_{e}, m_{\mu}, m_{\tau}) ~.
\label{hat_m_ell}
\end{eqnarray}
Throughout this work, we shall denote a diagonal matrix by putting a ``hat (~$\hat{}$~)'' on it, such
as the above $\hat m_{\ell}$.

The Yukawa terms ~$y_{\Sigma} (\overline{L_L} \Sigma)_{{\bf 1}} ~\tilde{\eta} + {\rm H.c.}$~ leads to
the neutrino Dirac mass and the corresponding charged lepton mass terms
\begin{eqnarray}
 {v_{\eta} \over \sqrt{2}} ~\overline{\nu_L} ~\hat Y_{\Sigma} ~N_{R}
 +v_{\eta} ~\overline{\ell_L} ~\hat Y_{\Sigma} ~E_R^- ~+ {\rm H.c.}~,
\label{y_Sigma_term}
\end{eqnarray}
after the $A_4$ singlet field ~$\eta$~ acquires the VEV ~$\langle \eta^{0} \rangle \equiv v_{\eta}$,~
which is assumed to be the electroweak scale: $v_{\eta} \sim v$.~
The Dirac mass matrix is given by
\begin{eqnarray}
 m_D = {v_{\eta} \over \sqrt{2}} ~ \hat Y_{\Sigma} = m_{\nu}^D ~I ~,
\label{mD}
\end{eqnarray}
where ~$m_{\nu}^D \equiv v_{\eta} y_{\Sigma} /\sqrt{2}$~ and the Yukawa coupling matrix
~$\hat Y_{\Sigma} \equiv y_{\Sigma} ~I$.~

The terms involving $M$ and $\chi$ give the mass terms of the right-handed Majorana neutrino $N_R$
and the heavy charged lepton $E_R^-$.
Taking the $A_{4}$ symmetry breaking scale to be above the electroweak scale, $i.e.$,
$\langle \chi_i \rangle > v$, one obtains the mass terms
\begin{eqnarray}
 {1 \over 2} ~\overline{(N_R)^c} ~M_N ~N_R +\overline{(E_R^+)^c} ~M_E ~E_R^- ~+ {\rm H.c.}~,
\label{heavy_Masses}
\end{eqnarray}
where the Majorana neutrino mass matrix $M_N$ and the heavy charged lepton mass matrix $M_E$ are given by
\begin{eqnarray}
 M_N = M_E =
 {\left(\begin{array}{ccc}
 M &  \lambda^{s}_{\chi} \upsilon_{\chi_{3}} &  \lambda^{s}_{\chi} \upsilon_{\chi_{2}} \\
 \lambda^{s}_{\chi} \upsilon_{\chi_{3}} &  M &  \lambda^{s}_{\chi} \upsilon_{\chi_{1}} \\
 \lambda^{s}_{\chi} \upsilon_{\chi_{2}} &  \lambda^{s}_{\chi} \upsilon_{\chi_{1}} &  M
 \end{array}\right)}~, ~
\label{MN_ME}
\end{eqnarray}
where $\langle \chi_{i} \rangle \equiv v_{\chi_{i}} ~(i=1,2,3)$. Both $M_N$ and $M_E$ are symmetric
matrices.  We note that there is no contribution to $M_N$ and $M_E$ from the ${\bf 3_a}$ term with
the coupling $\lambda_{\chi}^a$ in the Lagrangian.\footnote{See the details given in the subsection of
Appendix A.}
If the vacuum alignment of the $A_4$ triplet field $\chi$ is chosen to be
\begin{eqnarray}
 v_{\chi_1} \equiv v_{\chi} \neq 0 ~, ~~~~~ v_{\chi_2} = v_{\chi_3} = 0 ~,
\label{VEV_chi}
\end{eqnarray}
the matrices ~$M_N$~ and ~$M_E$~ become
\begin{eqnarray}
 M_N = M_E =
 M{\left(\begin{array}{ccc}
 1 &  0 &  0 \\
 0 &  1 &  \kappa ~e^{i \xi} \\
 0 &  \kappa ~e^{i \xi} &  1
 \end{array}\right)}~,
\label{MN_ME_2}
\end{eqnarray}
where ~$\kappa \equiv \left| \lambda_{\chi} v_{\chi} / M \right|$~ and the relative phase difference
~$\xi$~ is real.
The choice of VEV directions in Eq.~(\ref{VEV_chi}) and ~$\langle \varphi^0 \rangle$~ require a stable
alignment of the fields ~$\chi$ and $\Phi$, which is displayed in the Appendix.

For convenience, we change the basis for the SM charged lepton and heavy neutrino parts to be
diagonal as following:
\begin{eqnarray}
 && L_L ~\to~ L^{\rm d}_L \equiv (\nu^{\rm d}_L ~, ~\ell^{\rm d}_L)^T = U^{\dag}_{\omega} ~L_L
  \equiv (U^{\dag}_{\omega} ~\nu_L ~, ~U^{\dag}_{\omega} ~\ell_L)^T ~, ~~~~
  \ell_R ~\to~ \ell^{\rm d}_R = U_R^{\ell\dag} ~\ell_R = \ell_R ~,  \nonumber \\
 && N_R ~\to~ N^{\rm d}_R = (U_R^N)^{\dagger} N_R ~, ~~~~
  E_R^- ~\to~ E_R^{\rm d -} = (U_R^E)^{\dagger} E_R^- ~,
\label{basis}
\end{eqnarray}
where the diagonalization matrices ~$U_R^E = U_R^N$~ since ~$M_N = M_E$.~
Note that these states {\it with the superscript} ``d'' ($\nu^{\rm d}_L, ~\ell^{\rm d}_{L,R}$, etc)
are not yet final mass eigenstates, as can be seen below.
Then, in this basis the Yukawa interactions given in Eq.~(\ref{L_Yuk}) together with the charged gauge
interactions can be written in the form of the Type III seesaw Lagrangian
\begin{eqnarray}
 -{\cal L}
 &=& \overline{E^{\rm d}} ~\hat M_E ~E^{\rm d}
  + \Big\{ ~\frac{1}{2} ~\overline{(N^{\rm d}_R)^c} ~\hat M_N ~N^{\rm d}_R
  +\overline{\ell^{\rm d}_L} ~\hat m_\ell ~\ell^{\rm d}_R
  +\overline{\nu^{\rm d}_L} ~m'_{D} ~N^{\rm d}_R
  +\sqrt{2}~\overline{\ell^{\rm d}_L} ~m'_D ~E^{\rm d}_R  \nonumber \\
 && -g \Big[ ~\overline{E^{\rm d}_L}~ \gamma^{\mu} (N^{\rm d}_R)^c ~W_{\mu}^-
  +\overline{E^{\rm d}_R}~ \gamma^{\mu} N^{\rm d}_R ~ W_{\mu}^- \Big]
  -\frac{g}{\sqrt{2}} ~\overline{\ell^{\rm d}_L} \gamma^{\mu} \nu^{\rm d}_L ~W_{\mu}^-
  + ~{\rm H.c.} ~\Big\} ~,
\label{L_Yuk_gauge}
\end{eqnarray}
where the diagonal matrices ~$\hat M_E$~ and ~$\hat M_N$~ are given by
\begin{eqnarray}
 \hat M_E = \hat M_N = (U_R^N)^T ~M_N ~U_R^N = M ~{\rm diag}(a, 1, b) \equiv {\rm diag}(M_1, M_2, M_3)~,
\label{hatME}
\end{eqnarray}
with ~$a=\sqrt{1+\kappa^{2}+2\kappa\cos\xi}$~ and ~$b=\sqrt{1+\kappa^{2}-2\kappa\cos\xi}$.~
The diagonal elements for the heavy neutral and charged lepton mass matrices are $M_{1} = Ma$,
$M_{2} = M$ and $M_{3} = Mb$, which are real and positive.  For ~$\kappa \neq 0$,~ the diagonalization
matrix $U_R^N$ is
\begin{eqnarray}
  U_R^N = \frac{1}{\sqrt{2}} {\left(\begin{array}{ccc}
  0  &  \sqrt{2}  &  0 \\
  1 &  0  &  -1 \\
  1 &  0  &  1
  \end{array}\right)}{\left(\begin{array}{ccc}
  e^{i\frac{\alpha}{2}}  &  0  &  0 \\
  0  &  1  &  0 \\
  0  &  0  &  e^{i\frac{\beta}{2}}
  \end{array}\right)}~,
\label{URN}
\end{eqnarray}
with the phases
\begin{eqnarray}
 \alpha = \tan^{-1} \Big( \frac{-\kappa\sin\xi}{1+\kappa\cos\xi} \Big)
 ~~~{\rm and}~~~ \beta = \tan^{-1} \Big( \frac{-\kappa\sin\xi}{\kappa\cos\xi-1} \Big)~.
\label{alphs_beta}
\end{eqnarray}
In Eq.~(\ref{L_Yuk_gauge}), we have defined
~$E^{\rm d} = E^{\rm d}_R + E^{\rm d}_L \equiv E^{\rm d -}_R + (E^{\rm d +}_R)^c$,~ where
~$E^{\rm d}_{R, L} = P_{R, L} E^{\rm d}$~ with ~$P_{R, L} = (1 \pm \gamma_5)/2$,~
by using that ~$\hat M_E$~ is a real diagonal matrix.\footnote{If one defines
~$E = E_R + E_L \equiv E_R^- + (E_R^+)^c$~ in Eq.~(\ref{heavy_Masses}), it implies ~$M_E =
M_E^*$~ and ~$M_N = M_N^*$~ so that the phase ~$\xi$~ in Eq.~(\ref{MN_ME_2}) would vanish and Majorana
phases could not appear in the neutrino mass matrix.  However, this is not generally appropriate.}
The Dirac mass matrix ~$m'_{D}$ in Eq.~(\ref{L_Yuk_gauge}) is given by
\begin{eqnarray}
 m'_D  = \frac{v_{\eta}}{\sqrt{2}}U_{\omega}^{\dagger} ~\hat Y_{\Sigma} ~U_R^N
 = \frac{v_{\eta}y_{\Sigma}}{\sqrt{2}}{\left(\begin{array}{ccc}
  \frac{2}{\sqrt{6}} & \frac{1}{\sqrt{3}} &  0 \\
  -\frac{1}{\sqrt{6}} &  \frac{1}{\sqrt{3}}  &  -\frac{1}{\sqrt{2}} \\
  -\frac{1}{\sqrt{6}} &  \frac{1}{\sqrt{3}}  &  \frac{1}{\sqrt{2}}
 \end{array}\right)}
 {\left(\begin{array}{ccc}
  e^{i\frac{\alpha}{2}}  &  0  &  0 \\
  0  &  1  &  0 \\
  0  &  0  &  e^{i\frac{\beta-\pi}{2}}
 \end{array}\right)}~,
\label{Y_nu}
\end{eqnarray}
where $y_{\Sigma}$ is complex in general.  We note that the matrix product $U_{\omega}^{\dagger} ~U_R^N$
has the form of the so-called tri-bimaximal mixing matrix $U_{\rm TB}$ :
\begin{eqnarray}
 U_{\rm TB} = e^{i (\delta +\frac{\pi}{2})} {\left(\begin{array}{ccc}
  \frac{2}{\sqrt{6}} & \frac{1}{\sqrt{3}} &  0 \\
  -\frac{1}{\sqrt{6}} &  \frac{1}{\sqrt{3}}  &  -\frac{1}{\sqrt{2}} \\
  -\frac{1}{\sqrt{6}} &  \frac{1}{\sqrt{3}}  &  \frac{1}{\sqrt{2}}
 \end{array}\right)}
 {\left(\begin{array}{ccc}
  e^{i\frac{\alpha}{2}}  &  0  &  0 \\
  0  &  1  &  0 \\
  0  &  0  &  e^{i\frac{\beta-\pi}{2}}
 \end{array}\right)} ~,
\label{tribimaximal}
\end{eqnarray}
where ~$\delta$~ is an arbitrary phase. Here we have explicitly shown the possible Majorana phases
$\alpha$ and $(\beta -\pi)$, and the arbitrary phase $(\delta +\frac{\pi}{2})$~ in $U_{\rm TB}$.

Due to the existence of the mixing terms between $\nu^{\rm d}_L$ and $N^{\rm d}_R$ and between
$\ell^{\rm d}_L$ and $E^{\rm d}_R$, these states {\it with the superscript} ``d'' are not yet
final mass eigenstates. From Eq.~(\ref{L_Yuk_gauge}), the
lepton mass terms can be easily identified, such as the neutrino mass terms having the Type I seesaw
form
\begin{eqnarray}
 -{\cal L}_{\nu} = \frac{1}{2} ~\overline{{\cal N}_L} ~{\cal M}_{\nu}~ {\cal N}^{~c}_L
  +{\rm H.c.}~,~~~~~
 {\cal N}_{L} = {\left( \begin{array}{c}
  \nu^{\rm d}_{L} \\
  (N^{\rm d}_{R})^c
 \end{array}\right)}~,~~~
 {\cal M}_{\nu} = \left( \begin{array}{cc}
  0 &  m'_D \\
  m_D^{\prime ~T} & \hat M_N
 \end{array} \right) ~,
\label{nu_mass}
\end{eqnarray}
and the charged lepton mass terms
\begin{eqnarray}
 -{\cal L}_{\ell} = \overline{{\cal K}_{L}} ~{\cal M}_{\ell}~ {\cal K}_{R} +{\rm H.c.}~,~~~~~
 {\cal K}_{L,R} = {\left( \begin{array}{c}
  \ell^{\rm d}_{L,R} \\
  E^{\rm d}_{L,R}
 \end{array}\right)}~,~~~
 {\cal M}_{\ell} = \left( \begin{array}{cc}
  \hat m_{\ell} &  \sqrt{2}~ m'_D \\
  0 & \hat M_E
 \end{array} \right) ~.
\label{charged_mass}
\end{eqnarray}
Indeed, the full $6 \times 6$ mass matrices ${\cal M}_{\nu}$ and ${\cal M}_{\ell}$ are non-diagonal and
can be diagonalized by transforming the lepton fields from the states {\it with the superscript} ``d''
in Eq.~(\ref{L_Yuk_gauge}) to mass eigenstates which will be denoted by putting the {\it superscript}
``m'' as below:
\begin{eqnarray}
 {\cal N}_L ~\to~ {\cal N}^{\rm m}_L = U^{\dagger}~ {\cal N}_L ~,~~~~~~~~~~~~~~~
 {\cal K}_{L, R} ~\to~ {\cal K}^{\rm m}_{L, R} = X_{L, R}^{\dagger}~ {\cal K}_{L, R} ~,
\end{eqnarray}
where the lepton fields in the mass eigenstates are
\begin{eqnarray}
 {\cal N}^{\rm m}_L = {\left( \begin{array}{c}
  \nu^{\rm m}_{L} \\
  (N^{\rm m}_{R})^c
 \end{array}\right)}~,~~~~~~~~~~
 {\cal K}^{\rm m}_{L,R} = {\left( \begin{array}{c}
  \ell^{\rm m}_{L,R} \\
  E^{\rm m}_{L,R}
 \end{array}\right)}~,~~~
\end{eqnarray}
and the unitary matrices $U$ and $X_{L,R}$ can be written as
\begin{eqnarray}
 && U = {\left( \begin{array}{cc}
  U_{\nu\nu} & U_{\nu N} \\
  U_{N\nu} & U_{NN}
 \end{array}\right)}~, ~~~
 X_{L}={\left( \begin{array}{cc}
  X_{L\ell\ell} & X_{L\ell E} \\
  X_{LE\ell} & X_{LEE}
 \end{array}\right)}~, ~~~
 X_{R}={\left( \begin{array}{cc}
  X_{R\ell\ell} & X_{R\ell E} \\
  X_{RE\ell} & X_{REE}
 \end{array}\right)}~.
\label{6x6_diag}
\end{eqnarray}
Under the assumption ~$M \gg v_{\eta}, v$,~ up to order ~$(|y_{\Sigma}| v_{\eta} /M)^2$,~ we obtain
\begin{eqnarray}
 U_{\nu\nu} &=& \big( 1-U_{\nu N} U^{\dag}_{\nu N}/2 \big) ~U_{0}~,~~~~~~~~~~~~~~~~~~
  U_{\nu N}=m_{D} \hat M^{-1 T} \nonumber \\
 U_{N\nu} &=& -U^{\dag}_{\nu N} ~U_{0}~,~~~~~~~~~~~~~~~~~~~~~~~~~~~~~~~
  U_{NN} = 1 -U^{\dag}_{\nu N}U_{\nu N}/2 \nonumber\\
 X_{L\ell\ell} &=& \big( 1-m_{D} \hat M^{-1} \hat M^{-1 \dag} m^{\dag}_{D} \big) ~V_L^{\ell} ~,~~~~~~~~
  X_{L\ell E} = \sqrt{2} ~m_{D} \hat M^{-1} ~V_L^E \nonumber \\
 X_{LE\ell} &=& -\sqrt{2} ~\hat M^{-1 \dag} m^{\dag}_{D} ~V_L^{\ell} ~,~~~~~~~~~~~~~~~~~~
  X_{LEE} = \big( 1 -2 ~\hat M^{-1 \dag} m^{\dag}_{D} m_{D} \hat M^{-1} \big) ~V_L^E ~, \nonumber\\
 X_{R\ell\ell} &=& V_R^{\ell} ~,~~~~~~~~~~~~~~~~~~~~~~~~~~~~~~~~~~~~~~~
  X_{R\ell E} = \sqrt{2} ~\hat M^{-1} \hat M^{-1 \dag} m^{\dag}_{D} \hat m_{\ell} ~V_R^E \nonumber\\
 X_{RE\ell} &=& \sqrt{2} ~\hat m^{\dag}_{\ell} m_{D} \hat M^{-1} \hat M^{-1 \dag} ~V_R^{\ell} ~,~~~~~~~~~~
  X_{REE} = V_R^E ~,
\label{mixing_matrices}
\end{eqnarray}
where $\hat M \equiv \hat M_N = \hat M_E$.
$V_L^{\ell},~ V_R^{\ell},~ V_L^E$ and $V_R^E$ are the diagonalization matrices of the
hermitian matrices $\tilde m_{\ell} ~\tilde m^{\dag}_{\ell}$,
$\tilde m^{\dag}_{\ell} ~\tilde m_{\ell}$, $\tilde M_E ~\tilde M^{\dag}_E$ and
$\tilde M^{\dag}_E ~\tilde M_E$, respectively, which are expressed in Eq.~(\ref{A2}) of Appendix A.
For both light and heavy charged leptons, the next leading order terms in Eq.~(\ref{A2}) are negligibly
small, compared with the leading order terms, since ~$|M| \gg v_{\eta}$.
Especially, for the light charged leptons, the corrections to ~$\hat m_{\ell} ~\hat m^{\dag}_{\ell}$~
and ~$\hat m^{\dag}_{\ell} ~\hat m_{\ell}$~ first appear at order ~$(|y_{\Sigma}| v_{\eta} /M)^2$.

Up to order ~$|y_{\Sigma}| v_{\eta} /M$,~ the unitary matrix ~$U_{0}$~ in Eq.~(\ref{mixing_matrices})
is the diagonalization matrix of the $3 \times 3$ light neutrino mass matrix ~$m_{\nu}^{\rm mod}$:
\begin{eqnarray}
 U_0^T ~m_{\nu}^{\rm mod} ~U_0 = \hat m_{\nu}^{\rm mod} ~,
\label{mnu_diagonal}
\end{eqnarray}
where
\begin{eqnarray}
 m_{\nu}^{\rm mod}= -m'_D ~\hat M^{-1} ~m^{\prime ~T}_D ~, ~~~~~
 \hat m_{\nu}^{\rm mod} = {\rm diag}(m_1, m_2, m_3) ~,
\end{eqnarray}
with real and positive ~$m_i$~ $(i = 1, 2, 3)$.~
Due to Eq.~(\ref{Y_nu}),~Eq.~(\ref{mnu_diagonal}) holds if
\begin{eqnarray}
 U_0^* = U_{\omega}^{\dagger}~ U_R^N ~ e^{i (\frac{\pi}{2} +\delta)} = U_{\rm TB}
\label{U0_UTB}
\end{eqnarray}
and
\begin{eqnarray}
 \hat m_{\nu}^{\rm mod} = \frac{|y_{\Sigma}|^2~ v_{\eta}^2}{2} ~\hat M^{-1} ~,
\end{eqnarray}
where ~$y_{\Sigma} \equiv |y_{\Sigma}| ~e^{i \delta}$~ and the tri-bimaximal mixing matrix ~$U_{\rm TB}$~
is given in Eq.~(\ref{tribimaximal}).
In other words, the diagonalization matrix ~$U_0$~ naturally becomes the tri-bimaximal mixing matrix
~$U_{\rm TB}^*$.~
Therefore, with the relation (\ref{U0_UTB}),  Eq.~(\ref{mnu_diagonal}) can be rewritten as
\begin{eqnarray}
 m_{\nu}^{\rm mod} = U_{\rm TB} ~\hat m_{\nu}^{\rm mod} ~U_{\rm TB}^T ~,
\label{mnu_TB}
\end{eqnarray}
where the diagonal matrix ~$\hat m_{\nu}^{\rm mod}$~ is
\begin{eqnarray}
 \hat m_{\nu}^{\rm mod} = m_0 ~{\rm diag} \Big( ~\frac{1}{a}, ~1, ~\frac{1}{b} ~\Big) ~, ~~~~~
 {\rm with}~~ m_0 = \frac{|y_{\Sigma}|^2~ v_{\eta}^2}{2 M} ~,
\label{light_nu_mass}
\end{eqnarray}
Here $a$ and $b$ have been defined in Eq.~(\ref{hatME}).

It should be emphasized that being started from the Type III seesaw Lagrangian~(\ref{L_Yuk}) having
$A_4 \times Z_2$ symmetry, the tribimaximal mixing matrix $U_{\rm TB}$ is obtained in a natural way
as the diagonalization matrix of the light neutrino mass matrix, which is the
Pontecorvo-Maki-Nakagawa-Sakata (PMNS) matrix $U_{\rm PMNS}$ in the SM.
This feature is actually the same as in Type I seesaw case with $A_4$ flavor symmetry.

The above fact that the PMNS matrix naturally becomes the tribimaximal matrix ~$U_{\rm TB}$~ in this
model can be also shown directly from the charged gauge interactions as follows.
In the mass eigenstate basis the charged gauge interactions can be written as
\begin{eqnarray}
 {\cal L}_C &=& \frac{g}{\sqrt{2}} ~W^{-}_{\mu}
  ~\Big[ \big( \overline{\ell^{\rm m}_L}~ X^{\dag}_{L\ell\ell}
  +\overline{E^{\rm m}_L}~ X^{\dag}_{L\ell E} \big) ~\gamma^{\mu} ~\big( U_{\nu\nu} ~\nu^{\rm m}_{L}
  +U_{\nu N}~ (N^{\rm m}_R)^c \big)  \nonumber\\
 && ~+\sqrt{2} ~\big( \overline{\ell^{\rm m}_L}~ X^{\dag}_{LE\ell}
  +\overline{E^{\rm m}_L}~ X^{\dag}_{LEE} \big) ~\gamma^{\mu} ~\big( U_{N\nu} ~\nu^{\rm m}_{L}
  +U_{NN} (N^{\rm m}_R)^c \big) \nonumber\\
 && ~+\sqrt{2} ~\big( \overline{\ell^{\rm m}_R}~ X^{\dag}_{RE\ell}
  +\overline{E^{\rm m}_R}~ X^{\dag}_{REE})
  ~\gamma^{\mu} ~\big( U^{\ast}_{N\nu} ~(\nu^{\rm m}_L)^c + U^{\ast}_{NN} ~N^{\rm m}_{R}) \Big]
  +{\rm H.c.}~,
\label{charged_gauge}
\end{eqnarray}
which indicates the light lepton charged current
\begin{eqnarray}
 \frac{g}{\sqrt{2}} ~W^{-}_{\mu} ~\overline{\ell^{\rm m}_L}~ \gamma^{\mu}
   ~U_{\rm PMNS} ~\nu^{\rm m}_{L} +{\rm H.c.}
\label{charged_current}
\end{eqnarray}
with the PMNS matrix
\begin{eqnarray}
 U_{\rm PMNS} = X^{\dag}_{L\ell\ell} ~U_{\nu\nu} +\sqrt{2}~ X^{\dag}_{LE\ell} ~U_{N\nu}
  ~\simeq X^{\dag}_{L\ell\ell} ~U_{\nu\nu} ~.
\label{Type3_PMNS}
\end{eqnarray}
The approximation in~(\ref{Type3_PMNS}) is obvious from Eq.~(\ref{mixing_matrices}).
Since ~$U_{\nu\nu} \simeq U_{0}$~ and ~$X_{L\ell\ell} \simeq V_L^{\ell}$~ from
Eq.~(\ref{mixing_matrices}), and ~$V_L^{\ell} \simeq I$~ due to ~$\tilde m_{\ell} ~\tilde m^{\dag}_{\ell}
\simeq \hat m_{\ell} ~\hat m^{\dag}_{\ell}$~ from Eq.~(\ref{A2}), the PMNS matrix becomes
\begin{eqnarray}
 U_{\rm PMNS} \simeq V_L^{\ell \dag} ~U_{0} \simeq U_{\rm TB}^* ~.
\end{eqnarray}

\begin{table}[t]
\caption{\label{tab:data}Current best-fit values of $\theta_{12},\theta_{23},\Delta m^{2}_{\rm sol}$ and $\Delta m^{2}_{\rm atm}$ together with the $1 \sigma$ and $3 \sigma$ allowed
ranges of the neutrino oscillation parameters~\cite{valle}, and $\theta_{13}$ with a combined analysis of the data coming from T2K, MINOS, Double Chooz and Daya Bay experiments~\cite{Reactor, Machado:2011ar}.}
\begin{tabular}{|c|c|c|c|c|c|} \hline
 & $\Delta m^{2}_{\rm sol}/10^{-5}\mathrm{\ eV}^2$ & $\sin^2\theta_{12}$ &
 $\sin^{2}2\theta_{13}$ & $\sin^2\theta_{23}$ &
 $\Delta m^{2}_{\rm atm}/10^{-3}\mathrm{\ eV}^2$ \\ \hline \hline
Best-fit        &     7.59     &  0.312          &  0.089          &
0.52          &  2.50($-2.40$) \\ \hline
$1\sigma$ & $7.41 - 7.79$ & $0.295 - 0.329$  & $0.073 -
0.105$  & $0.45 - 0.58$(0.46-0.58)  & $2.34 - 2.59$-($2.48-2.31$) \\ \hline
$3\sigma$ & $7.09 - 8.19$ & $0.27 - 0.36$  & $0.042-0.136$
& $0.39 - 0.64$  & $2.14 - 2.76$$-(2.13-2.67)$ \\ \hline
\end{tabular}
\end{table}

Because of the observed hierarchy ~$|\Delta m^{2}_{\rm atm}| \equiv |\Delta m^{2}_{31}| \gg
\Delta m^{2}_{\rm sol} \equiv \Delta m^{2}_{21}>0$~ (as shown in Table~~\ref{tab:data}) and the
requirement of MSW resonance for solar neutrinos, from Eq.~(\ref{light_nu_mass}) there are two possible
neutrino mass hierarchies depending on the sign of ~$\cos\xi$~ (by definition, $\kappa > 0$)~: (i)
~$m_{1} < m_{2} < m_{3}$~ (normal hierarchy) corresponding to ~$\cos\xi > 0$~ and (ii)
~$m_{3} < m_{1} < m_{2}$~ (inverted hierarchy) corresponding to ~$\cos\xi < 0$. From
Eq.~(\ref{light_nu_mass}) the solar and atmospheric mass-squared differences are given by
\begin{eqnarray}
 \Delta m^{2}_{\rm sol}&\equiv& m^{2}_{2}-m^{2}_{1}=
  \frac{m^{2}_{0} ~\kappa (\kappa+2\cos\xi)}{1+\kappa^{2}+2\kappa\cos\xi}~,\nonumber\\
 \Delta m^{2}_{\rm atm}&\equiv& m^{2}_{3}-m^{2}_{1}=
  \frac{4m^{2}_{0} ~\kappa \cos\xi}{(1+\kappa^{2}-2\kappa\cos\xi)(1+\kappa^{2}+2\kappa\cos\xi)}~,
\label{deltam1}
\end{eqnarray}
which are constrained by the neutrino oscillation experimental results. Since the neutrino oscillation
data indicate that $\Delta m^{2}_{\rm Sol}$ is positive, we obtain the condition $\kappa > -2 \cos\xi$.
Also, from the data giving the value of the ratio of the mass-squared difference
~$R ~\equiv ~\Delta m^{2}_{\rm Sol}/\Delta m^{2}_{\rm Atm} \sim 3\times10^{-2}$,~we
find the other conditions ~$1 +\kappa^{2} \approx 2 \kappa \cos\xi$~ or ~$\kappa \approx -2 \cos\xi$.~
For the first case (corresponding to ~$M_{1,2} \gg M_{3}$)~ which implies ~$\cos\xi > 0$,~ the normal
hierarchy ~$m_{3}\gg m_{2}>m_{1}$~ is obtained.  By using the best-fit values of the neutrino oscillation
data for $R$,~ we find ~$\kappa \approx 0.75$~ or ~$1.24$~ for ~$\cos\xi \to 1$.
For the second case (corresponding to ~$M_{3}> M_{2} \gtrsim M_{1}$)~ which implies ~$\cos\xi < 0$,~
we find the inverted hierarchy ~$m_{2} \gtrsim m_{1}>m_{3}$.
From the best-fit values of the data for $R$,~ we have ~$\kappa \approx 2.01$~ for ~$\cos\xi \to -1$.

\section{Higher dimensional operators $-$ Deviation from Tri-bimaximal mixing}

The recent global fit analyses indicate that the mixing angle $\theta_{13}$ is non-zero at $1\sigma$ level.
In order to accommodate this fact in our framework, we introduce {\it higher dimensional operators which are also invariant under $SU(2)_L \times U(1)_Y \times A_4 \times Z_2$}, as before.
We assume that there is a cutoff scale $\Lambda$ above which there exists unknown physics.  Then below
the scale $\Lambda$, the higher dimensional operators express the effects from the unknown physics.

The effective dimension-five operators in the lepton sector, which are driven by the $\chi$-VEV alignment
and invariant under $SU(2)_L \times U(1)_Y \times A_4 \times Z_2$, can be expressed as
\begin{eqnarray}
 -{\cal L}_{\rm Yuk}^{~d=5} &=&
 \frac{y^{s}_{\chi}}{\Lambda} ~[(\overline{L_L} ~\Sigma)_{{\bf 3}_{s}} \cdot \chi]_{\bf 1}~ \tilde{\eta}
  +\frac{y^{a}_{\chi}}{\Lambda} ~[(\overline{L_L} ~\Sigma)_{{\bf 3}_{a}} \cdot \chi]_{\bf 1}~ \tilde{\eta}
 \nonumber\\
 &+& \frac{y^{s}_{e}}{\Lambda} ~[(\overline{L_L} ~\Phi)_{{\bf 3}_{s}} \cdot \chi]_{\bf 1}~ l_{R}
  +\frac{y^{s}_{\mu}}{\Lambda} ~[(\overline{L_L} ~\Phi)_{{\bf 3}_{s}} \cdot \chi]_{{\bf 1}'}~ l''_{R}
  +\frac{y^{s}_{\tau}}{\Lambda} ~[(\overline{L_L} ~\Phi)_{{\bf 3}_{s}} \cdot \chi]_{{\bf 1}''}~ l'_{R} \\
 &+& \frac{y^{a}_{e}}{\Lambda} ~[(\overline{L_L} ~\Phi)_{{\bf 3}_{a}} \cdot \chi]_{\bf 1}~ l_{R}
  +\frac{y^{a}_{\mu}}{\Lambda} ~[(\overline{L_L} ~\Phi)_{{\bf 3}_{a}} \cdot \chi]_{{\bf 1}'}~ l''_{R}
  +\frac{y^{a}_{\tau}}{\Lambda} ~[(\overline{L_L} ~\Phi)_{{\bf 3}_{a}} \cdot \chi]_{{\bf 1}''}~ l'_{R}
 + {\rm H.c.}~  \nonumber
\label{L_dim5}
\end{eqnarray}
Due to the above operators driven by the $\chi$ scalar field with VEV alignments in Eq.~(\ref{VEV_chi}),
the Dirac mass matrix in Eq.~(\ref{mD}) and the SM charged lepton mass matrix in Eq.~(\ref{m_ell})
are modified, while the heavy lepton masse matrices $M_N$ and $M_E$ are not affected.

After the electroweak symmetry breaking $\langle \eta^{0} \rangle = v_{\eta}$, the terms with the
couplings $y^{s,a}_{\chi}$ produce the off-diagonal elements of the Dirac mass matrix which can be
expressed as
\begin{eqnarray}
 {\left(\begin{array}{ccc}
 \overline{\nu_{L1}} &  \overline{\nu_{L2}} &  \overline{\nu_{L3}}
 \end{array}\right)}~ \frac{v_{\eta}}{\sqrt{2}} ~\Delta Y_{\Sigma}
 {\left(\begin{array}{c}
 N_{R1}  \\
 N_{R2}  \\
 N_{R3}
 \end{array}\right)} +{\rm H.c.} ,
\label{Y_nuN_term}
\end{eqnarray}
and
\begin{eqnarray}
 {\left(\begin{array}{ccc}
 \overline{\ell_{L1}} &  \overline{\ell_{L2}} &  \overline{\ell_{L3}}
 \end{array}\right)}~ v_{\eta} ~\Delta Y_{\Sigma}
 {\left(\begin{array}{c}
 E_{R1}  \\
 E_{R2}  \\
 E_{R3}
 \end{array}\right)} +{\rm H.c.},
\label{Y_lE_term}
\end{eqnarray}
where the deviation from the diagonal Yukawa matrix given in Eq.~(\ref{mD}), $\Delta Y_{\Sigma}$, is
given by
\begin{eqnarray}
 \Delta Y_{\Sigma} =
 {\left(\begin{array}{ccc}
   0 &  0 &  0 \\
   0 &  0 & (y^{s}_{\chi}+y^{a}_{\chi})\frac{\upsilon_{\chi}}{\Lambda} \\
   0 &  (y^{s}_{\chi}-y^{a}_{\chi})\frac{\upsilon_{\chi}}{\Lambda} & 0
 \end{array}\right)} = y_{\Sigma}
 {\left(\begin{array}{ccc}
   0 &  0 &  0 \\
   0 & 0 & y_{1}e^{i\rho_{1}} \\
   0 &  y_{2}e^{i\rho_{2}} & 0
 \end{array}\right)}~,
\end{eqnarray}
with $y_{1,2} = (|y^{s}_{\chi} \pm y^{a}_{\chi}| / |y_{\Sigma}|) (\upsilon_{\chi} /\Lambda)$ and
$\rho_{1,2} = {\rm arg}[(y^{s}_{\chi} \pm y^{a}_{\chi}) / y_{\Sigma}]$.

Similarly, the terms with the couplings ~$y^{s,a}_{e}$, $y^{s,a}_{\mu}$, $y^{s,a}_{\tau}$~ generate
corrections to the SM charged lepton mass matrix:
\begin{eqnarray}
 {\left(\begin{array}{ccc}
 \overline{\ell_{L1}} &  \overline{\ell_{L2}} &  \overline{\ell_{L3}}
 \end{array}\right)}~ v ~\Delta m_{\ell}
 {\left(\begin{array}{c}
   l_{R}  \\
   l''_{R}  \\
   l'_{R}
 \end{array}\right)} +{\rm H.c.} ~,
\label{m_ell_correc}
\end{eqnarray}
where the deviation from $m_{\ell}$ given in Eq.~(\ref{m_ell}), $\Delta m_{\ell}$, is given by
\begin{eqnarray}
 \Delta m_{\ell} =
 {\left(\begin{array}{ccc}
   0 &  0 &  0 \\
   (y^{s}_{e}+y^{a}_{e})\frac{\upsilon_{\chi}}{\Lambda} &
   (y^{s}_{\mu}+y^{a}_{\mu})\frac{\upsilon_{\chi}}{\Lambda} &
   (y^{s}_{\tau}+y^{a}_{\tau})\frac{\upsilon_{\chi}}{\Lambda} \\
   (y^{s}_{e}-y^{a}_{e})\frac{\upsilon_{\chi}}{\Lambda} &
   (y^{s}_{\mu}-y^{a}_{\mu})\frac{\upsilon_{\chi}}{\Lambda} &
   (y^{s}_{\tau}-y^{a}_{\tau})\frac{\upsilon_{\chi}}{\Lambda}
 \end{array}\right)} ~.
\end{eqnarray}
Combined with the previous mass matrix $m_{\ell}$, the {\it modified} SM charged
lepton mass matrix $m_{\ell}^{\rm mod}$ now becomes
\begin{eqnarray}
 m_{\ell}^{\rm mod} &=& m_{\ell} + \Delta m_{\ell}  \nonumber \\
 &=& U_{\omega}\sqrt{3}
 {\left(\begin{array}{ccc}
   m_{11} & m_{12} & m_{13} \\
   m_{21} & m_{22} & m_{23} \\
   m_{31} & m_{32} & m_{33}
 \end{array}\right)}
 \equiv U_{\omega} \tilde{U}_{L}^{\ell}~ {\rm diag}(m_{e},m_{\mu},m_{\tau})~
   (\tilde{U}_{R}^{\ell})^{\dag} ~,
\label{tilde_m_ell}
\end{eqnarray}
where
\begin{eqnarray}
 m_{11} &=& v (y_{e}+2f_{1}/3)~, ~~~ m_{12} = 2 v f_{2}/3~, ~~~~~~~~~~~~~~~~~~~
  m_{13} = 2 v f_{3}/3~,  \nonumber\\
 m_{21} &=& v (g_{1}-f_{1})/3~, ~~~~ m_{22} = v [y_{\mu}+(g_{2}-f_{2})/3]~, ~~~~
  m_{23} = v (g_{3}-f_{3})/3~,  \\
 m_{31} &=& -v (g_{1}+f_{1})/3~, ~~ m_{32} = -v (g_{2}+f_{2})/3~, ~~~~~~~~~
  m_{33} = v [y_{\tau}-(g_{3}+f_{3})/3]~, \nonumber
\end{eqnarray}
with ~$f_{1} = \upsilon_{\chi} y^{s}_{e}/\Lambda$,~ $f_{2} = \upsilon_{\chi} y^{s}_{\mu}/\Lambda$,~
$f_{3} = \upsilon_{\chi} y^{s}_{\tau}/\Lambda$,~ $g_{1} = -i \sqrt{3} \upsilon_{\chi} y^{a}_{e}/\Lambda$,~
$g_{2} = -i \sqrt{3} \upsilon_{\chi} y^{a}_{\mu}/\Lambda$,~
$g_{3} = -i \sqrt{3} \upsilon_{\chi} y^{a}_{\tau}/\Lambda$.
All $f_i$ and $g_i$ are in general complex.  The matrix $U_{\omega}$ is given in Eq.~(\ref{m_ell}).
Note that the diagonalization martix $\tilde{U}_R^{\ell}$ is not an identity matrix any more, which is
different from Eq.~(\ref{hat_m_ell}).
For the most natural case that the light charged lepton Yukawa couplings are hierarchical such as
$y_{\tau} \gg y_{\mu} \gg y_{e}$ and the corrected off-diagonal terms are smaller than the diagonal
ones in magnitude, we will make the following reasonable assumption
\begin{eqnarray}
 y_{\tau} \gg |f_{3}|, |g_{3}|\sim y_{\mu}\gg  |f_{2}|, |g_{2}|\sim y_{e}\gg  |f_{1}|, |g_{1}| ~,
\end{eqnarray}
or equivalently,
\begin{eqnarray}
 |m_{33}| \gg |m_{22}| \sim |m_{23}| \sim |m_{13}| \gg |m_{11}| \sim |m_{12}| \sim |m_{32}|\gg |m_{21}|
 \sim |m_{31}| ~.
\label{mass_hierarchy}
\end{eqnarray}
Under the above assumption, $\tilde{U}_L^{\ell}$ and $\tilde{U}_R^{\ell}$ can be obtained by
diagonalizing the matrices $U^{\dag}_{\omega} m_{\ell}^{\rm mod} m^{{\rm mod} \dag}_{\ell} U_{\omega}$
and $m^{{\rm mod} \dag}_{\ell} m_{\ell}^{\rm mod}$, respectively.
Notice that the mixing matrix $\tilde{U}_{L}^{\ell}$ becomes the part of the PMNS mixing matrix.
Owing to the strong hierarchy in Eq.~(\ref{mass_hierarchy}), $\tilde{U}_{L}^{\ell}$ can be approximated
as
\begin{eqnarray}
 \tilde{U}_{L}^{\ell} \simeq {\left(\begin{array}{ccc}
 1 & |\frac{m_{12}}{m_{22}}|e^{i\phi_{3}}  & |\frac{m_{13}}{m_{33}}|e^{i\phi_{2}}  \\
 -|\frac{m_{12}}{m_{22}}|e^{-i\phi_{3}} &  1 & |\frac{m_{23}}{m_{33}}|e^{i\phi_{1}} \\
 -|\frac{m_{13}}{m_{33}}|e^{-i\phi_{2}} &  -|\frac{m_{23}}{m_{33}}|e^{-i\phi_{1}} & 1
 \end{array}\right)} ~,
\label{tilde_U_L_ell}
\end{eqnarray}
where the phases $\phi_{i}~(i=1,2,3)$ are approximated as
\begin{eqnarray}
  \phi_{1} \simeq \frac{1}{2} \arg(m_{23}m^{\ast}_{33}) ~, ~~~~~
  \phi_{2} \simeq \frac{1}{2} \arg(m_{13}m^{\ast}_{33}) ~, ~~~~~
  \phi_{3} \simeq \frac{1}{2} \arg(m_{12}m^{\ast}_{22}) ~.
\end{eqnarray}

For convenience, let us change the basis for the SM charged lepton and heavy lepton (both neutral and
charged) parts to be diagonal:
\begin{eqnarray}
 && L_{L} \to (\tilde{U}^{\ell}_{L})^{\dag} U^{\dag}_{\omega} ~L_{L} ~,~~~~~
  \ell_{R} \to (\tilde{U}^{\ell}_{R})^{\dag} ~\ell_{R}~, \nonumber \\
 && N_{R} \to (U^N_{R})^{\dag} ~N_{R} ~,~~~~~~
  E_{R} \to (U^E_{R})^{\dag} ~E_{R} ~,~~~~~
  E_{L} \to (U^E_{R})^T ~E_{L} ~,
\end{eqnarray}
where $E_R \equiv E_R^-$ and $E_L \equiv (E_R^+)^c$, and the diagonalization matrices ~$U_R^E = U_R^N$~
due to ~$M_N = M_E$~ as in Eq.~(\ref{MN_ME}).~
Then the Yukawa and the charged gauge interactions have the same form as of Eq.~(\ref{L_Yuk_gauge})
with
\begin{eqnarray}
 \hat M_E &=& \hat M_N = (U_R^N)^T ~M_N~ U_R^N ~, ~~~~~
   \hat m_{\ell}^{\rm mod} = (\tilde{U}_L^{\ell})^{\dag} U^{\dag}_{\omega}
     ~m_{\ell}^{\rm mod}~ \tilde{U}_R^{\ell} ~,\nonumber\\
 m_D^{\prime {\rm mod}} &=&\frac{v_{\eta}}{\sqrt{2}}~(\tilde{U}_L^{\ell})^{\dag} U^{\dag}_{\omega}
   ~\hat Y_{\Sigma}~ U_R^N ~,
\end{eqnarray}
where $\hat M_E$, $\hat M_N$ and $\hat m_{\ell}^{\rm mod}$ are diagonal matrices, but in general
$m_D^{\prime {\rm mod}}$ is non-diagonal.
Because of the non-vanishing $m_D^{\prime {\rm mod}}$, the full $6 \times 6$ mass matrices ${\cal M}_{\nu}$
and ${\cal M}_{\ell}$, as defined in Eqs.~(\ref{nu_mass}) and (\ref{charged_mass}), are non-diagonal
with the Dirac mass matrix $m_D^{\prime {\rm mod}}$.
The $3 \times 3$ light neutrino mass matrix $m_{\nu}^{\rm mod}$ has the same form as of the Type I seesaw:
\begin{eqnarray}
 m_{\nu}^{\rm mod} &=& -m_D^{\prime {\rm mod}} ~\hat M_N^{-1}~ m^{\prime {\rm mod} ~T}_{D} \nonumber\\
 &=& -\frac{v_{\eta}^{2}}{2}~ [ (\tilde{U}_L^{\ell})^{\dag} U^{\dag}_{\omega} ~\hat Y_{\Sigma}~ U_R^N ]
   ~\hat M_N^{-1}~ [ (U_R^N)^T ~\hat Y_{\Sigma}^T~ U^*_{\omega} (\tilde{U}_L^{\ell})^* ] ~,
\label{mu_seesaw1}
\end{eqnarray}
which clearly shows that $m_{\nu}^{\rm mod}$ can not be diagonalized by the tri-bimaximal mixing matrix
$U_{\rm TB} = U_{\omega}^{\dagger}~ U_R^N ~ e^{i (\frac{\pi}{2} +\delta)}$,~ unlike the case shown in
Eq.~(\ref{mnu_TB}).  In other words, any matrix diagonalizing
$m_{\nu}^{\rm mod}$ should include a certain deviation from $U_{\rm TB}$.  The origin of the deviation from
~$U_{\rm TB}$~ is the corrections both to the Yukawa coupling matrix as shown in Eqs.~(\ref{Y_nuN_term})
and (\ref{Y_lE_term}), and to the SM charged lepton mass matrix as shown in Eq.~(\ref{m_ell_correc}).
In fact, the same feature can be obtained also in the Type I seesaw case with $A_4$ flavor symmetry, by
introducing the dimension-five operators similar to those shown in Eq.~(\ref{L_dim5}).
In the next section, we will investigate a new possibility that the above feature can be obtained through
pure Type III seesaw effects, which do not appear in the Type I seesaw case.

In order to explicitly show the deviation from the tri-bimaximal form, for simplicity, we assume that
the phase $\xi = 0$, defined in Eq.~(\ref{MN_ME_2}), which leads to the vanishing phases from heavy
lepton parts: $i.e.$, $\alpha=0$ and $\beta=0$ in Eq.~(\ref{alphs_beta}).
This assumption is equivalent to $\cos\xi = 1$ which corresponds to the normal hierarchy case for
the light neutrino masses in the previous section.
First, let us diagonalize ~$[U_{\omega} (\tilde{U}_L^{\ell}) ~m_{\nu}^{\rm mod}
~(\tilde{U}_L^{\ell})^T U^T_{\omega}]$,~ instead of ~$m_{\nu}^{\rm mod}$,~ by using a unitary matrix $V$:
\begin{eqnarray}
 V^{\dag} ~[U_{\omega} (\tilde{U}_L^{\ell}) ~m_{\nu}^{\rm mod} ~(\tilde{U}_L^{\ell})^T U^T_{\omega}]~ V^*
 &=& -\frac{v_{\eta}^{2}}{2}~ V^{\dag}
   ~[\hat Y_{\Sigma}~ U_R^N ~\hat M_N^{-1}~ (U_R^N)^T ~\hat Y_{\Sigma}^T] ~V^*  \nonumber \\
 &=& {\rm diag} (m_1^{\rm mod}, ~m_2^{\rm mod}, ~m_3^{\rm mod})  \\
 &=& \frac{y^2_\Sigma v^2_\eta}{2M}~ V^\dag
  {\left(\begin{array}{ccc}
   1 & 0 & 0 \\
   0 & A & G \\
   0 & G & B
   \end{array}\right)} V^* ~,  \nonumber
\end{eqnarray}
where $m_{i}^{\rm mod} ~(i=1,2,3)$ are the mass eigenvalues of the light neutrinos, and
 \begin{eqnarray}
  A &=& \frac{a(1-e^{i\rho_1}y_1)^2+b(1+e^{i\rho_1}y_1)^2}{2ab}~, \nonumber\\
  B &=& \frac{a(1-e^{i\rho_2}y_2)^2+b(1+e^{i\rho_2}y_2)^2}{2ab}~, \nonumber\\
  G &=& \frac{(1+e^{i\rho_1}y_1)(1+e^{i\rho_2}y_2)}{2a}-\frac{(1-e^{i\rho_1}y_1)(1-e^{i\rho_2}y_2)}{2b}~.
 \end{eqnarray}
Note that from the above expressions the PMNS matrix is given by
\begin{eqnarray}
U_{\rm PMNS} = (\tilde{U}_L^{\ell})^{\dag} U^{\dag}_{\omega}~ V ~.
\label{PMNS}
\end{eqnarray}
The diagonalization matrix $V$ is obtained as
\begin{eqnarray}
 V = e^{i\pi/2}~ {\left(\begin{array}{ccc}
 1 & 0 & 0 \\
 0 & e^{i\varphi_{1}}  &  0 \\
 0 & 0  &  e^{i\varphi_{2}}
 \end{array}\right)}
 {\left(\begin{array}{ccc}
 0 & 1 & 0 \\
 \cos\theta & 0  &  -\sin\theta \\
 \sin\theta & 0  &  \cos\theta
 \end{array}\right)}{\left(\begin{array}{ccc}
 e^{i\xi_{1}} & 0 & 0 \\
 0 & e^{i\xi_{2}}  &  0 \\
 0 & 0  &  e^{i\xi_{3}}
 \end{array}\right)} ~,
\label{V}
\end{eqnarray}
where the phases $\xi_{i}$ can be absorbed into the neutrino mass eigenstate fields, and the mixing
angle $\theta$ and the phase $\varphi_{21}$ are defined by
\begin{eqnarray}
 && \tan2\theta =\frac{2|AG^{\ast}+GB^{\ast}|}{|A|^{2}-|B|^{2}}~, \nonumber \\
 && \varphi_{21} \equiv \varphi_{2}-\varphi_{1} = \arg(GA^{\ast}+BG^{\ast})~.
\end{eqnarray}
It indicates that the angle $\theta$ and phase $\varphi_{21}$ go to $-\pi/4$ and $\pi$, respectively,
in the limit that $y_{1,2}$ vanish: $i.e.$, ~$\theta = -\pi/4 +\delta$~ with ~$|\delta| \ll 1$~ for
~$y_{1,2} \ll 1$.~
We will discuss below how the angle $\theta$ and phase $\varphi_{21}$ are correlated with the light
neutrino mixing angles and mass eigenvalues.
The light neutrino mass eigenvalues are given as
 \begin{eqnarray}
  (m^{\rm mod}_{1})^2 &=& m^{2}_{0} ~\Big( |A|^{2}\cos^{2}\theta+|B|^{2}\sin^{2}\theta+|G|^{2}
    +|AG^{\ast}+GB^{\ast}|\sin2\theta \Big)  \nonumber\\
  (m^{\rm mod}_{2})^2 &=& m^{2}_{0}  \nonumber\\
  (m^{\rm mod}_{3})^2 &=& m^{2}_{0} ~\Big( |A|^{2}\sin^{2}\theta+|B|^{2}\cos^{2}\theta+|G|^{2}
    -|AG^{\ast}+GB^{\ast}|\sin2\theta \Big) ~.
 \label{nu_mass2}
 \end{eqnarray}
Here the normal and inverted mass hierarchy cases correspond to $\theta=-\pi/4+\delta$ and
$\theta=\pi/4+\delta$, respectively.
The solar and atmospheric mass-squared differences are expressed as
\begin{eqnarray}
 \Delta m^{2}_{\rm sol}
  &=& m^{2}_{0} ~\Big( 1-|G|^{2}-|A|^{2}\cos^{2}\theta-|B|^{2}\sin^{2}\theta
   +|AG^{\ast}+GB^{\ast}|\sin2\theta \Big) ~,\nonumber\\
 \Delta m^{2}_{\rm atm}
  &=& -2m^{2}_{0} ~\frac{|AG^{\ast}+GB^{\ast}|}{\sin2\theta}~,
 \label{deltam2}
\end{eqnarray}
which are constrained by the neutrino oscillation experimental results given by Table~\ref{tab:data}.
Note that in the limit of $\theta\rightarrow-\pi/4$ and $\varphi_{21}\rightarrow\pi$ (equivalently
$y_{1,2}\rightarrow0$), as expected, Eq.~(\ref{deltam2}) turns back to Eq.~(\ref{deltam1}) for
$\xi = 0$ which corresponds to the normal mass hierarchy case.

In the followings, we will show that the non-zero $\theta_{13}$ can be generated in our $A_{4}$
symmetric model which leads to a certain deviation from the TBM through seesaw mechanism due to the
presence of the dimension-five operators driven by the $A_4$ triplet $\chi$ field.
In addition, we will show that the corrections through the SM charged lepton part can fit the $1\sigma$
experimental data.

\subsection{With negligible corrections from the SM charged lepton sector: ~$\tilde{U}_{L}^{\ell} = I$ }

In the case of $\tilde{U}_{L}^{\ell} = I$, from Eqs.~(\ref{m_ell}) and (\ref{V}), the lepton mixing
matrix $U_{\rm PMNS}$ can be written as
\begin{eqnarray}
 U_{\rm PMNS} = U^{\dag}_{\omega} V
 = e^{i\pi/2} \frac{1}{\sqrt{3}} ~{\left(\begin{array}{ccc}
  c e^{i\varphi_{1}} +s e^{i\varphi_{2}} & ~~1~~ &  c e^{i\varphi_{2}} -s e^{i\varphi_{1}}  \\
 -c e^{i(\varphi_{1} +\frac{\pi}{3})} -s e^{i(\varphi_{2}-\frac{\pi}{3})} & ~~1~~ &
  s e^{i(\varphi_{1} +\frac{\pi}{3})} -c e^{i(\varphi_{2} -\frac{\pi}{3})}  \\
 -c e^{i(\varphi_{1} -\frac{\pi}{3})} -s e^{i(\varphi_{2} +\frac{\pi}{3})} & ~~1~~ &
  s e^{i(\varphi_{1} -\frac{\pi}{3})} -c e^{i(\varphi_{2} +\frac{\pi}{3})}
 \end{array}\right)}~,
\label{PMNS1}
\end{eqnarray}
where $s \equiv \sin\theta$ and $c \equiv\cos\theta$.  The common phase $e^{i\pi/2}$ has no physical
meaning so that it can be neglected. It is clear that in the limit of $y_{1,2}=0$ (equivalently,
~$\theta=-\pi/4$ and $\varphi_{21}=\pi$ for the normal mass hierarchy case)~the exact TBM is
restored in Eq.~(\ref{PMNS1}).  By transformations $e\rightarrow e e^{i\alpha_{1}}$,
$\mu\rightarrow \mu e^{i\beta_{1}}$, $\tau\rightarrow \tau e^{i\beta_{2}}$ and
$\nu_{2}\rightarrow\nu_{2}e^{i(\alpha_{1}-\alpha_{2})}$, Eq.~(\ref{PMNS1}) can be rewritten as
 \begin{eqnarray}
  U_{\rm PMNS}=
 {\left(\begin{array}{ccc}
 |U_{e1}| & |U_{e2}| & U_{e3}e^{-i\alpha_{1}} \\
 U_{\mu1}e^{-i\beta_{1}} & U_{\mu2}e^{i(\alpha_{1}-\alpha_{2}-\beta_{1})} &  |U_{\mu3}| \\
 U_{\tau1}e^{-i\beta_{2}} & U_{\tau2}e^{i(\alpha_{1}-\alpha_{2}-\beta_{2})} & |U_{\tau3}|
 \end{array}\right)}~,
 \label{PMNS3}
 \end{eqnarray}
where $\alpha_{i}=\arg(U_{ei})$~ ($i=1,2,3$), ~$\beta_{1}=\arg(U_{\mu3})$ and
$\beta_{2}=\arg(U_{\tau3})$, and $U_{\zeta j}$ is an element of the PMNS matrix, with
$\zeta = e,\mu,\tau$ corresponding to the lepton flavors and $j=1,2,3$ corresponding to the light
neutrino mass eigenstates. Each elements of $U_{\rm PMNS}$ in Eq.~(\ref{PMNS3}) can be related to the
conventional parameters of the PMNS matrix~\cite{pdg}. Then, the reactor angle $\theta_{13}$
is written as
\begin{eqnarray}
 \sin\theta_{13} &=& |U_{e3}|=\frac{1}{\sqrt{3}}\sqrt{1-\sin2\theta \cos\varphi_{21}}~.
\label{angle13}
\end{eqnarray}
Using the $3\sigma~(1\sigma)$ experimental bounds on $|U_{e3}|$, we obtain the bounds on
$\sin2\theta\cos\varphi_{21}$:
~$0.89 \lesssim \sin2\theta \cos\varphi_{21} \leq 0.94$~
$(0.92 \lesssim \sin2\theta \cos\varphi_{21} \leq 0.97)$.
As will be shown below, these bounds are more stringent than that from $\theta_{12}$.

\begin{figure*}[t]
\begin{tabular}{c}
\includegraphics[width=6.5cm]{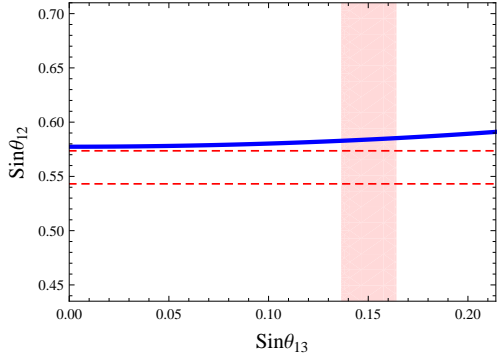}
\end{tabular}
\caption{\label{Fig0} Plot of $\sin\theta_{12}$ in Eq.~(\ref{angle1213}) versus
$\sin\theta_{13}$. Here the horizontal dashed lines represent $1\sigma$ experimental
bounds on $\sin\theta_{12}$ shown in Table~\ref{tab:data}.
The red band shows $1\sigma$ experimental bound in Eq.~(\ref{expdata})
}
\end{figure*}

The solar and atmospheric neutrino mixings are governed by
\begin{eqnarray}
 \sin^{2}\theta_{12} &=& \frac{|U_{e2}|^{2}}{1-|U_{e3}|^{2}}
  = \frac{1}{2+\sin2\theta \cos\varphi_{21}}~,\nonumber\\
 \sin^{2}\theta_{23} &=& \frac{|U_{\mu3}|^{2}}{1-|U_{e3}|^{2}}
  = \frac{1-\sin2\theta \cos(2\pi/3-\varphi_{21})}{2+\sin2\theta \cos\varphi_{21}}~.
\label{angle12}
\end{eqnarray}
The above relations indicate that $\sin^{2}\theta_{12} = 1/3$ and $\sin^{2}\theta_{23} = 1/2$ in
the limit of $\theta = \pi/4$ and $\varphi_{21} = 0$, and a deviation from those values of the
mixing angles are strongly constrained by $\theta$ and $\varphi_{21}$.
Using the $3\sigma$ experimental bound on the solar mixing angle, we obtain the constraint:
$0.78 \lesssim \sin2\theta \cos\varphi_{21} \leq 1$.
Combining this constraint with the expression in Eq.~(\ref{angle12}) leads to $\sin^{2}\theta_{12}
\geq 1/3$, which is disfavored by the $1\sigma$ experimental upper bound:
$\sin^{2}\theta_{12} = 0.331 <1/3$.
On the other hand, from Eqs.~(\ref{angle13}) and (\ref{angle12}) we obtain
a correlation between the solar mixing angle $\theta_{12}$ and the reactor mixing one $\theta_{13}$:
\begin{eqnarray}
 \sin\theta_{12}=\frac{1}{\sqrt{3}}\frac{1}{\sqrt{1-\sin^{2}\theta_{13}}}\geq\frac{1}{\sqrt{3}}~.
\label{angle1213}
\end{eqnarray}
Fig.~\ref{Fig0} displays this correlation between $\theta_{12}$ and $\theta_{13}$, and shows the lower
bound of the solar mixing angle $\sin\theta_{12}\geq1/\sqrt{3}$.

In the next section, in comparison with the above results, we shall discuss the phenomenological
consequences of the case that contributions from the SM charged lepton sector are sizable.
In the case that the $1\sigma$ experimental bound is taken seriously into account, this discussion
shall be also interesting.

\subsection{With sizable corrections from the SM charged lepton sector }

The diagonalization matrix $\tilde{U}_{L}^{\ell}$ of the SM charged lepton mass matrix can modify the
PMNS matrix to be consistent with $1\sigma$ experimental data shown in Table~\ref{tab:data}, by
generating sizable effects.  The modified lepton mixing matrix can be written as
\begin{eqnarray}
 U_{\rm PMNS} &=& (\tilde{U}^{\ell}_{L})^{\dag} U^{\dag}_{\omega} V  \nonumber\\
 &=&{\left(\begin{array}{ccc}
 U^v_{11}-U^{\ell}_{12}U^v_{21}-U^{\ell}_{13}U^v_{31} ~&~
  \frac{1}{\sqrt{3}}-\frac{1}{\sqrt{3}}(U^{\ell}_{12}+U^{\ell}_{13}) ~&~
  U^v_{13}-U^{\ell}_{12}U^v_{23}-U^{\ell}_{13}U^v_{33} \\
 U^v_{21}+U^{\ell\ast}_{12}U^v_{11}-U^{\ell}_{23}U^v_{31} &
  \frac{1}{\sqrt{3}}-\frac{1}{\sqrt{3}}(U^{\ell}_{23}-U^{\ell\ast}_{12})  &
  U^v_{23}-U^{\ell}_{23}U^v_{33}+U^{\ell\ast}_{12}U^v_{13} \\
 U^v_{31}+U^{\ell\ast}_{13}U^v_{11}+U^{\ell\ast}_{23}U^v_{21} &
  \frac{1}{\sqrt{3}}+\frac{1}{\sqrt{3}}(U^{\ell\ast}_{13}+U^{\ell\ast}_{23}) &
  U^v_{33}+U^{\ell\ast}_{13}U^v_{13}+U^{\ell\ast}_{23}U^v_{23}
 \end{array}\right)}~,
\label{PMNS2}
\end{eqnarray}
where $U^v_{ij}$ is an element of the matrix $U^{\dag}_{\omega} V$ given in Eq.~(\ref{PMNS1}), and
$U^{\ell}_{ij}$ is an element of $\tilde{U}_{L}^{\ell}$ given in Eq.~(\ref{tilde_U_L_ell}).
With the same manipulation as in Eq.~(\ref{PMNS3}), the reactor angle $\theta_{13}$ and solar mixing angle
$\theta_{12}$ can be expressed as
\begin{eqnarray}
 \sin\theta_{13}
 &=& |U^v_{13}+U^{\ell}_{12}U^v_{23}+U^{\ell}_{13}U^v_{33}| ~
  \simeq \frac{1}{\sqrt{3}} \sqrt{1-\sin2\theta\cos\varphi_{21}+\epsilon\lambda}~, \nonumber\\
 \sin^{2}\theta_{12}
 &=& \frac{1}{3} \frac{|1-U^{\ell}_{12}-U^{\ell}_{13}|^{2}}
  {1 -|U^v_{13}-U^{\ell}_{12}U^v_{23}-U^{\ell}_{13}U^v_{33}|^{2}}
  \simeq \frac{1-2\epsilon\cos\phi_{3}}{2+\sin2\theta\cos\varphi_{21}-\epsilon\lambda}~,
\label{theta13}
\end{eqnarray}
where
\begin{eqnarray}
 \lambda = \cos\phi_{3}+\sqrt{3}\sin\phi_{3}\cos2\theta
  -\sin2\theta ~[\cos(\varphi_{21}-\phi_{3}-\pi/3)+\cos(\varphi_{21}+\phi_{3}-\pi/3)] ~,
\end{eqnarray}
and we have assumed
\begin{eqnarray}
 \epsilon \equiv |U^{\ell}_{12}| \gg |U^{\ell}_{13}|\approx|U^{\ell}_{23}| ~.
\label{assumption}
\end{eqnarray}
\begin{figure*}[t]
\begin{tabular}{cc}
\includegraphics[width=6.5cm]{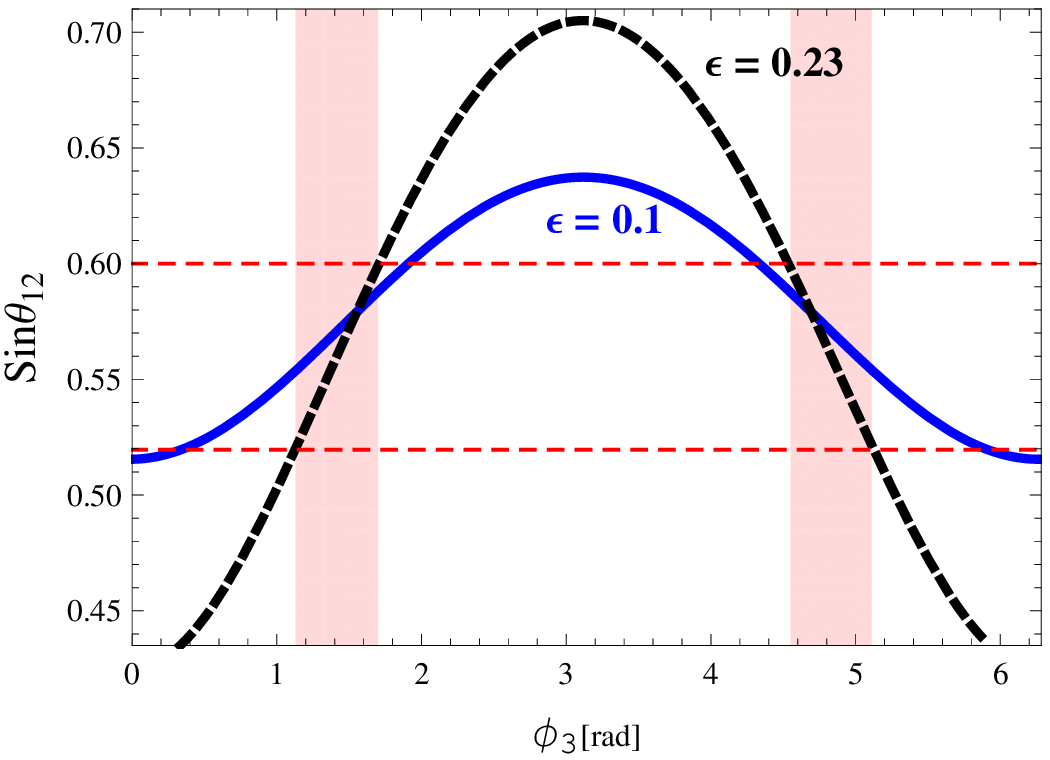}&
\includegraphics[width=6.5cm]{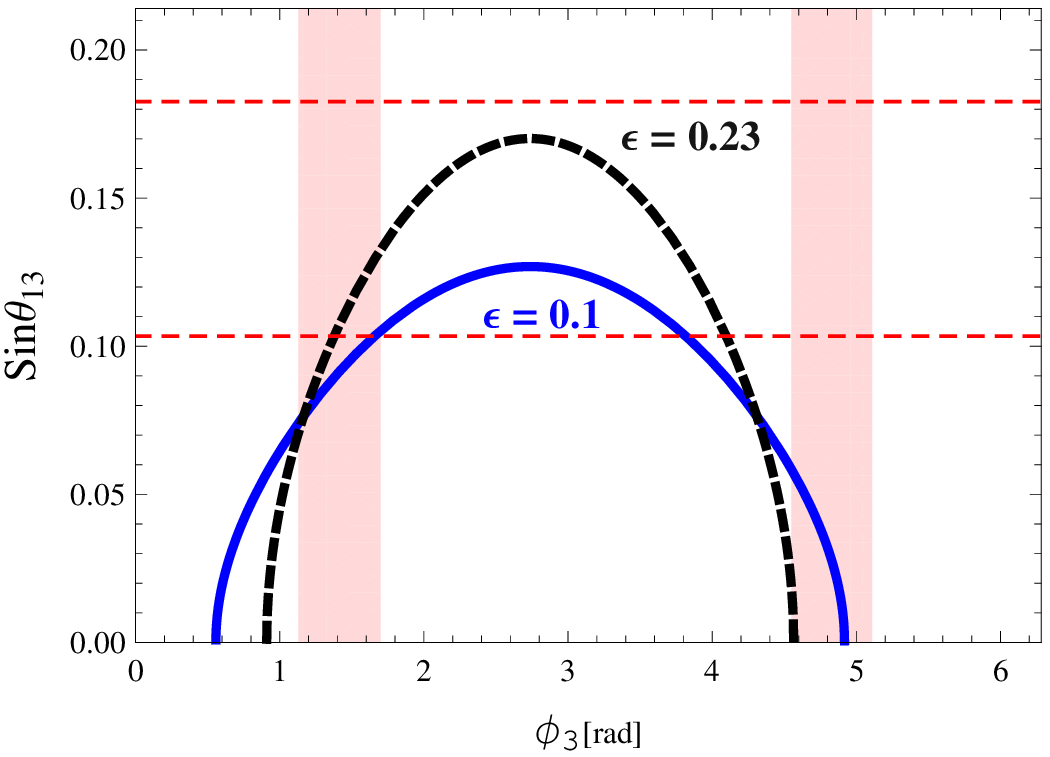}
\end{tabular}
\caption{\label{Figa} Plots of $\sin\theta_{12}$ and $\sin\theta_{13}$ as a function of $\phi_{3}$ [rad].
The (blue) solid and (black) dashed lines correspond to $\epsilon = 0.1$ and $\epsilon = 0.23$,
respectively, for $\theta=-43^{\circ}$ and $\varphi_{21}=183^{\circ}$.
The red bands are allowed regions for $\phi_{3}$ which is constrained by $\sin\theta_{12}$.
Here the horizontal dashed lines represent $3\sigma$ experimental bounds in Table~\ref{tab:data}.}
\end{figure*}
\mbox{} By comparing Eq.~(\ref{theta13}) with (\ref{angle13}) and (\ref{angle12}), it is clearly seen that the amount of the
modification effects to $\theta_{12}$ and $\theta_{13}$ depends on the parameters $\epsilon$ and
$\lambda$. For example, Fig.~\ref{Figa} shows how the solar mixing angle $\theta_{12}$ and reactor mixing angle $\theta_{13}$ depend
on the parameters $\phi_{3}$ and $\epsilon$ for fixed values of $\theta$ and $\varphi_{21}$:
the (blue) solid and (black) dashed lines correspond to $\epsilon=0.1$ and $\epsilon=0.23$
\footnote{The value $\epsilon=0.23$ corresponds to sine of the Cabibbo angle.}, respectively,
for $\theta=-43^{\circ}$ and $\varphi_{21}=183^{\circ}$ which are chosen to be values a little deviated
from $\theta = -45^{\circ}$ and $\varphi_{21} = 180^{\circ}$ equivalent to $y_{1,2} = 0$.
As can be seen in the left plot on $\theta_{12}$ of Fig.~\ref{Figa}, for $\epsilon=0.23$, there are two allowed regions on the phase $\phi_{3}$, that is, $1.1\lesssim\phi_{3}{\rm[rad]}\lesssim1.7$ and $4.6\lesssim\phi_{3}{\rm[rad]}\lesssim5.1$.  The right plot on $\theta_{13}$ of Fig.~\ref{Figa} shows that the measured value of $\theta_{13}$ favors only one region, $1.1\lesssim\phi_{3}{\rm[rad]}\lesssim1.7$.

\begin{figure*}[t]
\begin{tabular}{c}
\includegraphics[width=6.5cm]{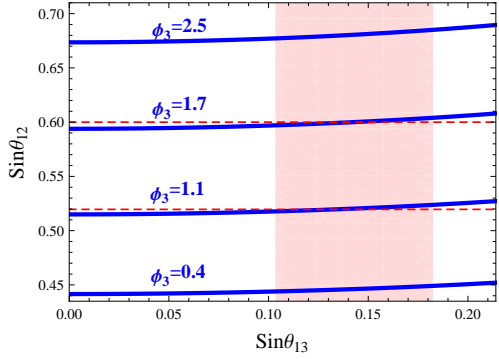}
\end{tabular}
\caption{\label{Figb} Plot of $\sin\theta_{12}$ versus $\sin\theta_{13}$ with the
varying phase $\phi_{3}$ [rad] for $\epsilon=0.23$. Here the horizontal dashed lines and red band
represent $3\sigma$ experimental bounds of $\theta_{12}$ and $\theta_{13}$, respectively, in Table \ref{tab:data}.}
\end{figure*}

Similar to Eq.~(\ref{angle1213}), from Eq.~(\ref{theta13}) we find a correlation
modified by the SM charged lepton sector between the solar mixing angle $\theta_{12}$ and the
reactor mixing one $\theta_{13}$:
\begin{eqnarray}
 \sin\theta_{12}=\sqrt{\frac{1-2\epsilon\cos\phi_{3}}{3(1-\sin^{2}\theta_{13})}}~.
\label{theta1213}
\end{eqnarray}
In comparison with Eq.~(\ref{angle1213}), the solar mixing angle in Eq.~(\ref{theta1213}) can be
sizably changed by the parameters $\epsilon$ and $\phi_{3}$.  Fig.~\ref{Figb} shows a correlation
between $\theta_{12}$ and $\theta_{13}$ for $\epsilon=0.23$, where the solid lines correspond to
$\phi_{3}=2.5,~1.7,~1.1,~0.4$ [rad] from the bottom, respectively.
For a fixed value $\epsilon=0.23$, there is a region of $\phi_{3}$, {\it i.e.} $1.1\lesssim\phi_{3}{\rm[rad]}\lesssim1.7$, satisfying the experimental data of $\theta_{12}$ and $\theta_{13}$ at $3\sigma$.
Comparing Fig.~\ref{Figb} to Fig.~\ref{Fig0}, we see that the value of $\sin\theta_{12}$ can vary to
a large extent, depending on $\phi_3$ which arises from the SM charged lepton effects.

\begin{figure*}[t]
\begin{tabular}{cc}
\includegraphics[width=6.5cm]{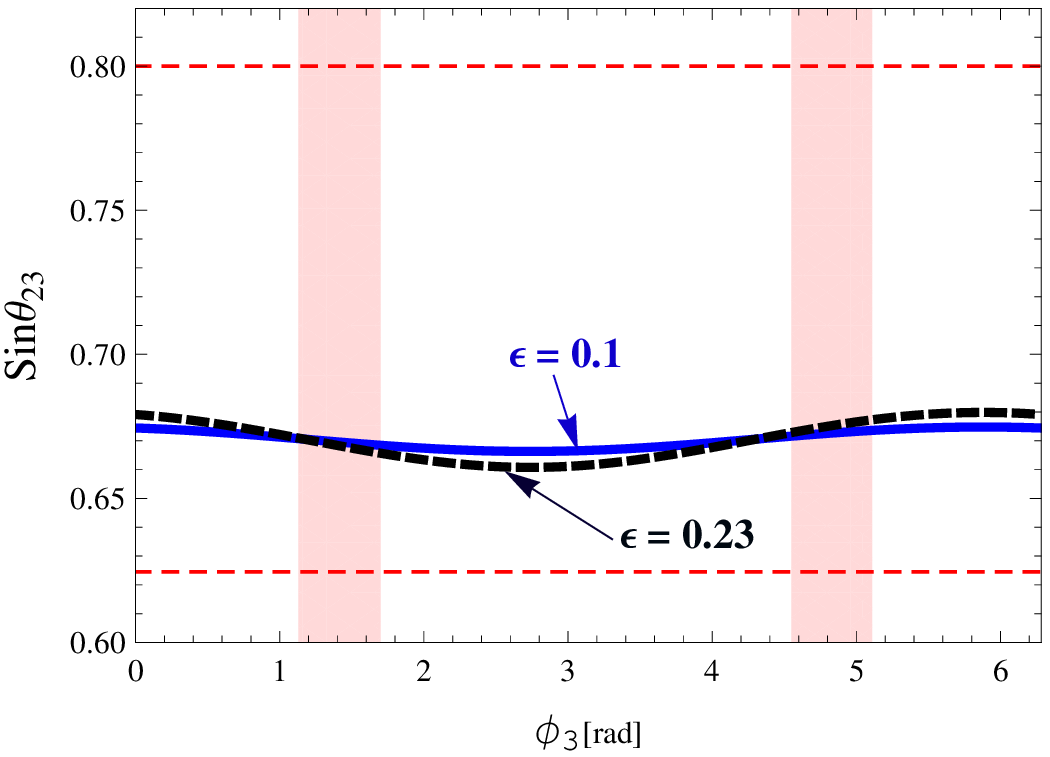}&
\includegraphics[width=6.5cm]{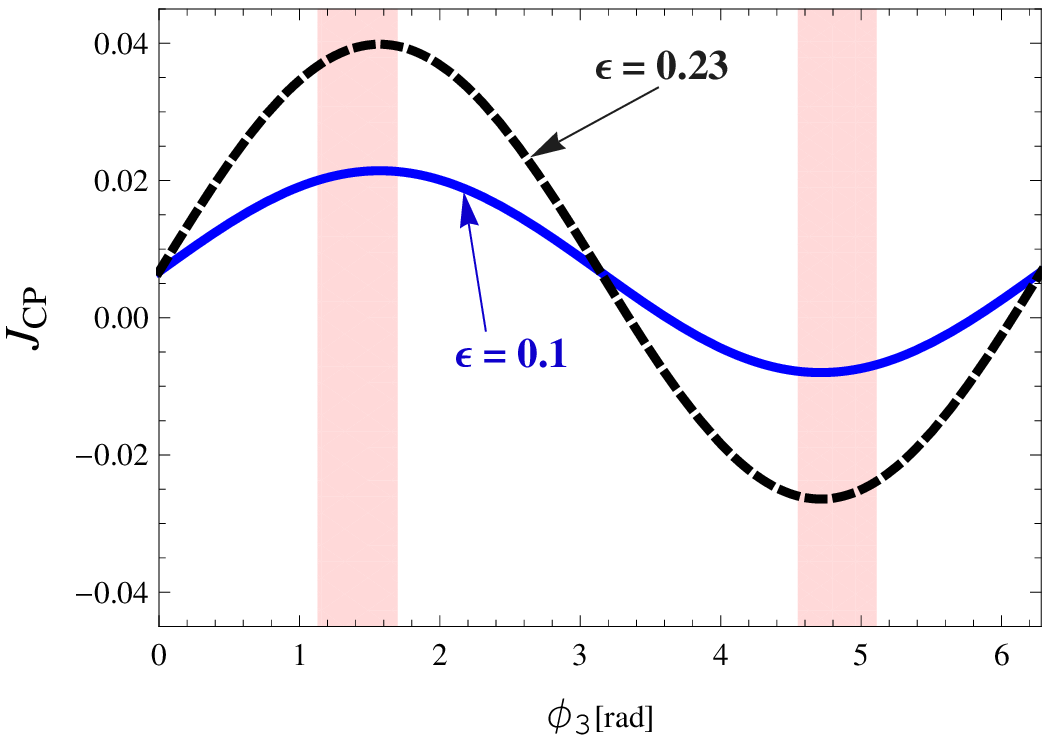}
\end{tabular}
\caption{\label{Figc} Plots of $\sin\theta_{23}$ (left) and $J_{CP}$ (right) as
a function of $\phi_{3}$ [rad]. In both plots, the (blue) solid and (black) dashed lines
correspond to $\epsilon=0.1$ and $\epsilon=0.23$, respectively, for $\theta=-43^{\circ}$
and $\varphi_{21}=183^{\circ}$.  Here the horizontal dashed lines represent $3\sigma$
experimental bounds of $\theta_{23}$ in Table \ref{tab:data}. The red bands come from the constraint of the experimental data $\theta_{12}$.}
\end{figure*}

Also, the atmospheric mixing angle can be modified as
\begin{eqnarray}
 \sin^{2}\theta_{23}
 &=& \frac{|U^v_{23}-U^{\ell}_{23}U^v_{33}+U^{\ell\ast}_{12} U^v_{13}|^{2}}
  {1-|U^v_{13}-U^{\ell}_{12}U^v_{23}-U^{\ell}_{13}U^v_{33}|^{2}}
  \simeq \frac{1-\sin2\theta\cos(2\pi/3-\varphi_{21})-\epsilon\lambda}
   {2+\sin2\theta\cos\varphi_{21}-\epsilon\lambda} ~,
\label{angle2}
\end{eqnarray}
where we have used Eq.~(\ref{assumption}). Again, the amount of the modification effects to $\theta_{12}$ and $\theta_{23}$ depends on the
parameters $\epsilon$ and $\lambda$.
It is very interesting to note that for $\sin2\theta \approx -1$ and $\cos\phi_{3}\simeq -1$, we have
~$\lambda \approx 0$, in which case the angle $\theta_{23}$ is not much modified from that in
Eq.~(\ref{angle12}), but only $\theta_{12}$ can be modified sizably by the SM charged lepton part.
For an illustration, we show plots of $\sin\theta_{12}$ in the left plot of Fig.~\ref{Figa} and $\sin\theta_{23}$ in the left plot of Fig.~\ref{Figc} as a function of $\phi_{3}$[rad], respectively: in both plots, the (blue) solid and (black)
dashed lines correspond to $\epsilon=0.1$ and $\epsilon=0.23$, respectively, for $\theta=-43^{\circ}$
and $\varphi_{21}=183^{\circ}$. Here the horizontal dotted lines represent $3\sigma$ experimental
bounds in Table \ref{tab:data}. And the red bands come from the constraint of experimental data of $\theta_{12}$.
We see that the value of $\sin\theta_{12}$ is sensitive to $\phi_3$ and $\epsilon$, while
the value of $\sin\theta_{23}$ varies relatively small.  In particular, from Figs.~\ref{Figa} and
\ref{Figc}, one can see the deviations of $\sin\theta_{12}$, $\sin\theta_{23}$ and $\sin\theta_{13}$
from their TBM values of $1 /\sqrt{3}$, $1 /\sqrt{2}$ and 0, respectively, depending on $\phi_3$ and
$\epsilon$.
It is also obvious from these two figures that there are allowed values of $\phi_3$ and $\epsilon$
to satisfy the $3\sigma$ experimental bounds on $\sin\theta_{12}$, $\sin\theta_{23}$ and
$\sin\theta_{13}$: e.g., $\phi_3 \sim 1.5$ [rad] for both $\epsilon = 0.1$ and 0.23.

Interestingly enough, CP violating phases arise from the dimension-five operators driven by the $\chi$
field and they are directly related to the low energy Dirac CP phase which can be measured, in
principle, in long baseline neutrino oscillation experiments~\cite{Freund:1999gy}.
By using the conventional parametrization of the PMNS matrix~\cite{pdg} and Eq.~(\ref{PMNS2}) one can deduce a expression for Dirac CP phase $\delta_{CP}$ which can be written as
\begin{eqnarray}
  \delta_{CP}
 &=& -\arg \left(\frac{\frac{U^{\ast}_{e1}U_{e3}U_{\tau1}U^{\ast}_{\tau3}}{c_{12}c^{2}_{13}c_{23}s_{13}}+c_{12}c_{23}s_{13}}{s_{12}s_{23}}\right)~.
\label{angle2}
\end{eqnarray}
Equivalently, the strength of the low energy CP violation measurable through neutrino oscillation
defined by Jarlskog invariant, $J_{CP}={\rm Im}[U_{e1}U_{\mu2}U^{\ast}_{e2}U^{\ast}_{\mu1}]$, could be expressed roughly in terms of our parameters
\begin{eqnarray}
  J_{CP}
 &\simeq& \frac{\sqrt{3}}{18}\left(\cos2\theta+\epsilon\sin2\theta(\sqrt{3}\cos\varphi_{21}-\sin\varphi_{21})\sin\phi_{3}\right)~.
\label{Jcp}
\end{eqnarray}
The right plot of Fig.~\ref{Figc} shows the behavior of the $J_{CP}$ as a function of $\phi_{3}$. As pointed out in Fig.~\ref{Figa}, the measured value of $\theta_{13}$ favors the region $1.1\lesssim\phi_{3}{\rm[rad]}\lesssim1.7$, which in turn means that $J_{CP}$ has a non-vanishing value, indicating a signal of maximal $CP$ violation.

It is worth noting that the features discussed above can be similarly obtained also in the Type I seesaw case
with the $A_4 \times Z_2$ symmetry.  In the Type III seesaw case, because of their origin from the same $SU(2)_L$ triplet, the
heavy neutral ($N$) and charged ($E$) leptons appear in the Lagrangian usually on the same footing, as shown in the previous
and this section. It is thus unlikely in the Type III seesaw with the $A_4 \times Z_2$ symmetry to find sizable effects from only
either $N$ or $E$ to the charged lepton mass terms or the neutrino Dirac mass terms.
However, the presence of the heavy charged lepton ($E$) in the Type III case leads to unique physical consequences differentiating
from those of the Type I case, such as decays of $E$ (through the gauge interactions given in Eq.~(\ref{charged_gauge})) and new
tree level FCNC processes, which can be tested in future experiments.

\section{Conclusion}

The seesaw mechanism is a promising way to explain the tiny masses of neutrinos, but it cannot provide
a solution for the puzzling pattern of mixing among different lepton flavors.  An interesting approach
for understanding the pattern of the mixing matrix in the lepton sector is to invoke certain family
symmetries which constrain the flavor structure of couplings of Yukawa interactions.

Motivated by the recent neutrino data from Daya Bay and RENO Collaborations,
we have studied the phenomenology of neutrino mixing angles in the Type III seesaw model with $A_4$ flavor
symmetry. Stating with the leptonic Yukawa interactions having a $SU(2)_L \times U(1)_Y \times A_4
\times Z_2$ symmetry which is spontaneously broken at a scale much higher than the EW scale, we have
shown that at tree level the TBM form of the lepton mixing PMNS matrix can be obtained in a natural
way.  From the current neutrino experimental data, either normal or inverted hierarchical case of
neutrino masses is allowed, depending on the sign of a particular parameter in our analysis.

By introducing higher dimensional operators, we have explicitly shown that the lepton mixing matrix
generally has a deviation from the TBM form such that it can explain the non-zero mixing angle
$\theta_{13}$ indicated by recent experimental data.
With negligible corrections from the charged lepton sector to the lepton mixing matrix, our result is
consistent with all the neutrino experimental bounds, such as $\Delta m_{\rm sol}^2$,
$\Delta m_{\rm atm}^2$, $\sin^2 \theta_{12}$, $\sin^2 \theta_{23}$ and $|U_{e3}|$ at $3 \sigma$ level,
but our prediction for the possible value of $\sin^2 \theta_{12}$ is disfavored by the data at
$1 \sigma$ level.
In the presence of effective dimension-5 operators driven by $SU(2)_{L}\times U(1)_{Y}$ singlet scalar fields we have found that sizable contributions from the charged lepton part modify the lepton mixing matrix
with which all the neutrino data can be accommodated through phase effects.
We have shown that although two regions on the phase $\phi_{3}$, $1.1 \lesssim\phi_{3}{\rm[rad]}\lesssim 1.7$ and $4.6 \lesssim\phi_{3}{\rm[rad]}\lesssim 5.1$, are allowed by the experimental data of $\theta_{12}$, the measured value of $\theta_{13}$ favors the former. In particular, the recently measured best-fit value of $\theta_{13} = 8.68^{\circ}$
can be understood in our framework in a consistent way with the constraints from the other mixing
angles $\theta_{12}$ and $\theta_{23}$. Furthermore, we have found that the leptonic $CP$ violation characterized by the Jarlskog invariant has a non-vanishing value, indicating a signal of maximal $CP$ violation $J_{CP}\simeq0.04$, which could be tested in the future experiments such as the upcoming long baseline neutrino
oscillation ones.

\acknowledgments{
We thank Xiao-Gang He for helpful discussions.
The work of C.S.K. was supported in part by the National Research Foundation of Korea (NRF)
grant funded by Korea government of the Ministry of Education, Science and Technology (MEST)
(No. 2011-0027275), (No. 2012-0005690) and (No. 2011-0020333).
}

\appendix
\section{}
\subsection{Comments on Eq.~(\ref{MN_ME})}

The term ${\rm Tr}[(\overline{\tilde \Sigma^{c}}\Sigma)_{{\bf 3}_{a}}]$ would lead to the terms
$[\overline{(N_R)^c} ~N_R]_{{\bf 3}_a}$ and $[\overline{(E_R^+)^c} ~E_R^-]_{{\bf 3}_a}$
$+[\overline{(E_R^-)^c} ~E_R^+]_{{\bf 3}_a}$.  But, the right-handed Majorana neutrino term
$[\overline{(N_R)^c} ~N_R]_{{\bf 3}_a}$ identically vanishes due to the property of a Majorana particle.
In contrast, for the heavy charged leptons, after $A_4$ symmetry breaking, the ${\bf 3_a}$ term leads to
$\overline{(E_R^+)^c} ~M_E^a ~E_R^-+~\overline{(E_R^-)^c}~ M_E^a ~E_R^+
= \overline{(E_R^+)^c} ~(M_E^a + M_E^{a T}) ~E_R^- = 0$, where
\begin{eqnarray}
 M_E^a =
 {\left(\begin{array}{ccc}
 0 &  \lambda^{a}_{\chi} \upsilon_{\chi_{3}} & \lambda^{a}_{\chi} \upsilon_{\chi_{2}} \\
 -\lambda^{a}_{\chi} \upsilon_{\chi_{3}} & 0 & \lambda^{a}_{\chi} \upsilon_{\chi_{1}} \\
 -\lambda^{a}_{\chi} \upsilon_{\chi_{2}} & -\lambda^{a}_{\chi} \upsilon_{\chi_{1}} & 0
 \end{array}\right)} = - M_E^{a T} ~.
\label{A1}
\end{eqnarray}

\subsection{Comments on Eq.~(\ref{mixing_matrices})}

The hermitian matrices ~$\tilde m_{\ell} ~\tilde m^{\dag}_{\ell}$~,
~$\tilde m^{\dag}_{\ell} ~\tilde m_{\ell}$~, ~$\tilde M_E ~\tilde M^{\dag}_E$~ and
~$\tilde M^{\dag}_E ~\tilde M_E$,~ respectively, which are given by~\cite{Bandyopadhyay:2009xa}
\begin{eqnarray}
 \tilde m_{\ell} ~\tilde m^{\dag}_{\ell} &=& \hat m_{\ell} ~\hat m^{\dag}_{\ell}
 -\Big( m'_D ~\hat M^{-1} ~\hat M^{-1 \ast} ~m'^{~\dag}_D ~\hat m_{\ell} ~\hat m^{\dag}_{\ell}
 + {\rm H.c.} \Big) ~,\nonumber\\
 \tilde m^{\dag}_{\ell} ~\tilde m_{\ell} &=& \hat m^{\dag}_{\ell} ~\hat m_{\ell}
  -4 ~\hat m^{\dag}_{\ell} ~m'_D ~\hat M^{-1} ~\hat M^{-1 \dag} ~m'^{~\dag}_D ~\hat m_{\ell}~,\nonumber\\
 \tilde M_E \tilde M^{\dag}_E &=& \hat M ~\hat M^{\dag}
  + \Big( \hat M ~m'^{~\dag}_D ~m'_D ~\hat M^{-1} + {\rm H.c.} \Big)
  + 2 \hat M^{-1 \dag} ~m'^{~\dag}_D ~\hat m_{\ell} ~\hat m^{\dag}_{\ell} ~m'_D ~\hat M^{-1} \nonumber\\
  && +\hat M^{-1 \dag} ~m'^{~\dag}_D ~m'_D ~m'^{~\dag}_D ~m'_D ~\hat M^{-1} + ... \nonumber\\
 \tilde M^{\dag}_E \tilde M_E &=& ~\hat M^{\dag} ~\hat M +2 m'^{~\dag}_D ~m'_D
  +\Big( 2 ~\hat M^{-1} ~\hat M^{-1 \dag} ~m'^{~\dag}_D ~\hat m_{\ell} ~\hat m^{\dag}_{\ell} ~m'_D
  + {\rm H.c.} \Big) + ...
\label{A2}
\end{eqnarray}
\section{Higgs Potential and vacuum alignments discussed in Section II}
\begin{figure}[h]
\epsfig{figure=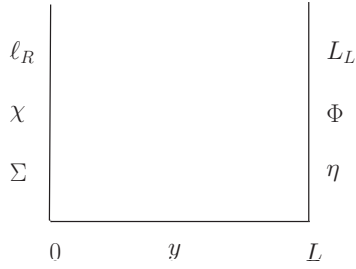,width=5cm,angle=0}
\caption{\label{fig:exd}
The fifth dimension and locations of scalar and fermion fields.}
\end{figure}

We are going to briefly discuss these vacuum alignments discussed in Sec. II, because it is
nontrivial to ensure that the different vacuum alignments of $\langle \varphi^{0} \rangle =
(\upsilon,\upsilon,\upsilon)$, $\langle \eta^{0} \rangle = v_{\eta} ~(\sim v)$ and
$\langle \chi \rangle=(\upsilon_{\chi},0,0)$ in Eq.~(\ref{VEV_chi}) are preserved.
There is a generic way to prohibit the problematic interaction terms by physically separating $\chi$
and $(\Phi,\eta)$.
Here we solve the vacuum alignment problem by extending the model with a spacial extra dimension
$y$~\cite{Altarelli:2005yp}. We assume that each field lives on the 4D brane either at $y = 0$ or at
$y = L$, as shown in Fig.~\ref{fig:exd}. The heavy neutrino masses arise from local operators at $y=0$,
while the charged fermion masses and the neutrino Yukawa interactions are realized by non-local effects
involving both branes.
A detailed explanation of this possibility is beyond the scope of this paper.

Then, the most general renormalizable scalar potentials of $\Phi, \eta$ and $\chi$, invariant under
$SU(2)_{L}\times U(1)_{Y}\times A_{4}\times Z_{2}$, are given by
\begin{eqnarray}
V_{y=L} &=& \mu^{2}_{\Phi}(\Phi^{\dag}\Phi)_{\mathbf{1}}
 +\lambda^{\Phi}_{1}(\Phi^{\dag}\Phi)_{\mathbf{1}}(\Phi^{\dag}\Phi)_{\mathbf{1}}
 +\lambda^{\Phi}_{2}(\Phi^{\dag}\Phi)_{\mathbf{1^\prime}}(\Phi^{\dag}\Phi)_{\mathbf{1^{\prime\prime}}}
 +\lambda^{\Phi}_{3}(\Phi^{\dag}\Phi)_{\mathbf{3}_{s}}(\Phi^{\dag}\Phi)_{\mathbf{3}_{s}}\nonumber\\
  &+&\lambda^{\Phi}_{4}(\Phi^{\dag}\Phi)_{\mathbf{3}_{a}}(\Phi^{\dag}\Phi)_{\mathbf{3}_{a}}
  +i\lambda^{\Phi}_{5}(\Phi^{\dag}\Phi)_{\mathbf{3}_{s}}(\Phi^{\dag}\Phi)_{\mathbf{3}_{a}}
  +\mu^{2}_{\eta}(\eta^{\dag}\eta)+\lambda^{\eta}(\eta^{\dag}\eta)^{2}\nonumber\\
  &+&\lambda^{\Phi\eta}_{1}(\Phi^{\dag}\Phi)_{\mathbf{1}}(\eta^{\dag}\eta)
  +\lambda^{\Phi\eta}_{2}(\Phi^{\dag}\eta)(\eta^{\dag}\Phi)
  +\lambda^{\Phi\eta}_{3}(\Phi^{\dag}\eta)(\Phi^{\dag}\eta)
  +\lambda^{\Phi\eta\ast}_{3}(\eta^{\dag}\Phi)(\eta^{\dag}\Phi) ~,
\label{potential1}\\
V_{y=0}  &=&\mu^{2}_{\chi}(\chi\chi)_{\mathbf{1}}
 +\lambda^{\chi}_{1}(\chi\chi)_{\mathbf{1}}(\chi\chi)_{\mathbf{1}}
  +\lambda^{\chi}_{2}(\chi\chi)_{\mathbf{1}^\prime}(\chi\chi)_{\mathbf{1}^{\prime\prime}}
  +\lambda^{\chi}_{3}(\chi\chi)_{\mathbf{3}}(\chi\chi)_{\mathbf{3}}
  +\xi^{\chi}(\chi\chi\chi)_{\mathbf{1}} ~,
\label{potential2}
\end{eqnarray}
where $\mu_{\Phi},\mu_{\eta},\mu_{\chi}$ and $\xi^{\chi}$ are of the mass dimension 1, while
$\lambda^{\Phi}_{1,...,5}$, $\lambda^{\eta}$, $\lambda^{\chi}_{1,...,3}$ and
$\lambda^{\Phi\eta}_{1,...,3}$ are all dimensionless.
From Eqs.~(\ref{potential1}) and (\ref{potential2}), it is easy to check that the vacuum stabilities
of global minima are guaranteed.

The minimum condition of the potential $V_{y=0}$ is
 \begin{eqnarray}
  \left. \frac{\partial V_{y=0}}{\partial \chi_1} \right|_{\langle \chi_1 \rangle = v_{\chi}}
  = 2v_{\chi} \Big[ \mu^2_{\chi} + 2(\lambda_1^{\chi}
    +\lambda_2^{\chi})v^2_{\chi} \Big] = 0~,
 \end{eqnarray}
and $\left. \frac{\partial V_{y=0}}{\partial \chi_{2,3}}\right|_{\langle \chi_{2,3} \rangle = 0} = 0$
are automatically satisfied.
On the other hand, the minimum conditions for the potential on the brane $y = L$ are
 \begin{eqnarray}
  \left. \frac{\partial V_{y=L}}{\partial \varphi^0_i} \right|_{\langle \varphi^0_i \rangle,
    \langle \eta \rangle}
  &=& 2v \Big[ \mu^2_{\Phi} + 2(3\lambda_1^{\Phi} + 2\lambda_3^{\Phi})v^2+(\lambda^{\Phi\eta}_{1}
    +\lambda^{\Phi\eta}_{2}+\lambda^{\Phi\eta}_{3}+\lambda^{\Phi\eta\ast}_{3})v^{2}_{\eta} \Big]
    = 0 ~,\nonumber\\
  \left. \frac{\partial V_{y=L}}{\partial \eta} \right|_{\langle \varphi^0_i \rangle,
    \langle \eta \rangle}
  &=& 2v_\eta \Big[ \mu^2_{\eta} + 2\lambda^{\eta}v^{2}_{\eta} +(\lambda^{\Phi\eta}_{1}
    +\lambda^{\Phi\eta}_{2} +\lambda^{\Phi\eta}_{3}+\lambda^{\Phi\eta\ast}_{3})v^2 \Big] = 0 ~,
 \end{eqnarray}
where $\langle \varphi^0_i \rangle = v ~ (i = 1,2,3)$ and $\langle \eta \rangle = v_{\eta}$ are used.
We obtain three independent equations for the three unknowns $v$, $v_{\eta}$ and $v_{\chi}$.
Thus the configurations needed in our scenario can be realized at tree level.
The stability of these vacuum alignments under higher order corrections is not explored in this work.


\end{document}